\documentclass[twocolumn,preprintnumbers,aps,pra,superscriptaddress,amsmath,amssymb]{revtex4-1}

\usepackage{graphicx}
\usepackage{dcolumn}
\usepackage{bm}
\usepackage{bbm}
\usepackage{hyperref}

\begin{document} 
\title{Scanning Nano-spin Ensemble Microscope for Nanoscale Magnetic and Thermal Imaging}

\author{Jean-Philippe~Tetienne}\email{jtetienne@unimelb.edu.au}
\affiliation{Centre for Quantum Computation and Communication Technology, The University of Melbourne, Melbourne Victoria 3010, Australia}
\affiliation{Bio21 Institute and School of Chemistry, The University of Melbourne, Melbourne Victoria 3010, Australia}

\author{Alain~Lombard}
\affiliation{Centre for Quantum Computation and Communication Technology, The University of Melbourne, Melbourne Victoria 3010, Australia}
\affiliation{D{\'e}partement de Physique, Ecole Normale Sup{\'e}rieure de Cachan, 94235 Cachan, France}

\author{David~A.~Simpson}
\affiliation{School of Physics, The University of Melbourne, Melbourne Victoria 3010, Australia}

\author{Cameron~Ritchie}
\affiliation{Bio21 Institute and School of Chemistry, The University of Melbourne, Melbourne Victoria 3010, Australia}

\author{Jianing~Lu}
\affiliation{Bio21 Institute and School of Chemistry, The University of Melbourne, Melbourne Victoria 3010, Australia}

\author{Paul~Mulvaney}
\affiliation{Bio21 Institute and School of Chemistry, The University of Melbourne, Melbourne Victoria 3010, Australia}

\author{Lloyd~C.~L.~Hollenberg}
\affiliation{Centre for Quantum Computation and Communication Technology, The University of Melbourne, Melbourne Victoria 3010, Australia}
\affiliation{Bio21 Institute and School of Chemistry, The University of Melbourne, Melbourne Victoria 3010, Australia}
\affiliation{School of Physics, The University of Melbourne, Melbourne Victoria 3010, Australia}

\date{\today}

\begin{abstract}

Quantum sensors based on solid-state spins provide tremendous opportunities in a wide range of fields from basic physics and chemistry to biomedical imaging. However, integrating them into a scanning probe microscope to enable practical, nanoscale quantum imaging is a highly challenging task. Recently, the use of single spins in diamond in conjunction with atomic force microscopy techniques has allowed significant progress towards this goal, but generalisation of this approach has so far been impeded by long acquisition times or by the absence of simultaneous topographic information. Here we report on a scanning quantum probe microscope which solves both issues, by employing a nano-spin ensemble hosted in a nanodiamond. This approach provides up to an order of magnitude gain in acquisition time, whilst preserving sub-100 nm spatial resolution both for the quantum sensor and topographic images. We demonstrate two applications of this microscope. We first image nanoscale clusters of maghemite particles through both spin resonance spectroscopy and spin relaxometry, under ambient conditions. Our images reveal fast magnetic field fluctuations in addition to a static component, indicating the presence of both superparamagnetic and ferromagnetic particles. We next demonstrate a new imaging modality where the nano-spin ensemble is used as a thermometer. We use this technique to map the photo-induced heating generated by laser irradiation of a single gold nanoparticle in a fluid environment. This work paves the way towards new applications of quantum probe microscopy such as thermal/magnetic imaging of operating microelectronic devices and magnetic detection of ion channels in cell membranes.   

\end{abstract}

\maketitle

In recent years, quantum sensing has attracted increasing attention because it opens up the possibilities for non-invasive sensing with nanoscale mapping potential and single atom sensitivity. The model system for these studies has been the negatively charged nitrogen-vacancy (NV) centre in diamond, which has shown great promise as an atomic-sized sensor operating under ambient conditions \cite{Hall2013,Rondin2014,Schirhagl2014}. The NV centre is a photostable point defect whose electronic ground state is a spin triplet that can be initialised and read out by optical means \cite{Gruber1997,Doherty2013}. This makes it possible to detect the spin resonance of a single NV centre and coherently control its spin state. Owing to interaction with the local environment, measurement of the spectral and temporal properties of an NV centre's spin provides access to a number of local properties, such as the magnetic field, electric field, pressure or temperature \cite{Maze2008,Gopi2008,Dolde2011,Doherty2014,Acosta2010}. Furthermore, NV sensors are capable not only of measuring static fields but also fluctuating ones. This, in turn, allows retrieval of spectral information from a sample \cite{Cole2009,Hall2009,Meriles2010,deLange2011,McGuinness2013,Schafer2014}. This capability was recently used to perform nuclear magnetic resonance spectroscopy on a small ensemble of protons \cite{Mamin2013,Staudacher2013}.   

\begin{figure*}[bt!]
\begin{center}
\includegraphics[width=1\textwidth]{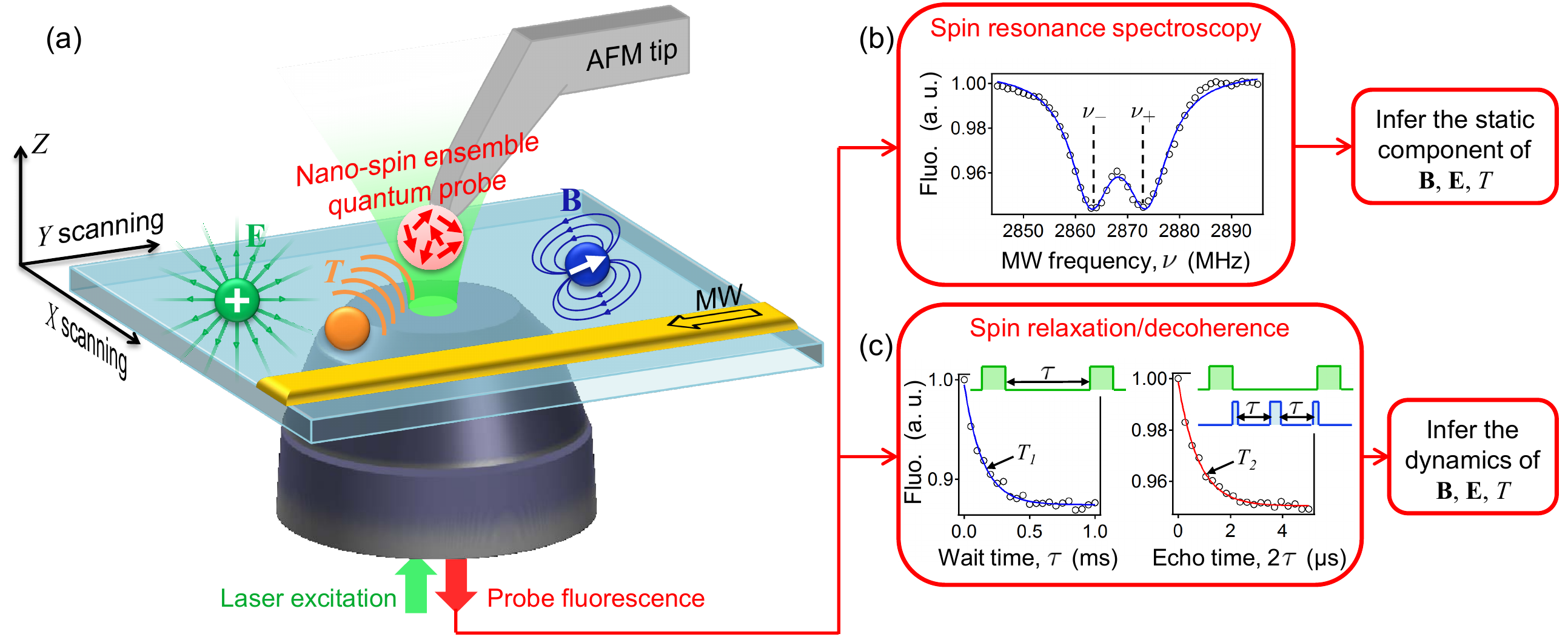}
\caption{Schematic of the scanning nano-spin ensemble microscope. (a) The probe consists of a nanodiamond containing a small ensemble of electronic spins (symbolised by the red arrows) grafted onto the tip of an atomic force microscope (AFM). Optical excitation and readout, combined with microwave (MW) excitation, enable quantum measurements of the spin ensemble properties such as spin resonance, relaxation or decoherence. Scanning the sample relative to the probe produces images of the sample with magnetic ({\bf B} field), electric ({\bf E} field) or temperature ($T$) contrast, depending on the probing technique used. (b) Optically detected spin resonance spectrum of a nanodiamond on a tip. The solid line is the fit to a sum of two Lorentzian functions centred at frequencies $\nu_\pm=D\pm E$, providing the zero-field splitting parameters $D=2868.3\pm0.1$ MHz and $E=5.4\pm0.1$ MHz. (c) Spin relaxation (left) and spin decoherence (right) curve of a nanodiamond on the AFM tip. The inset depicts the sequence of laser (green) and MW (blue) pulses employed. Solid lines are fits to a single exponential decay, revealing a spin relaxation time $T_1=142\pm9~\mu$s and spin coherence time $T_2=780\pm30$ ns.} 
\label{Fig1}
\end{center}
\end{figure*}

In order to obtain spatial information about the sample's properties, the NV probe must be scanned relative to the sample. This is commonly achieved by using an atomic force microscope (AFM), which enables a probe to be scanned relative to a sample, while precisely controlling the probe-sample distance. Integrating an NV probe into an AFM can be achieved by grafting a nanodiamond onto the tip of the AFM \cite{Rondin2012,Tisler2013,Schell2014,Tetienne2014}, or by fabricating an all-diamond AFM tip hosting an NV centre \cite{Maletinsky2012,Luan2015}. An alternative route is to make use of an NV centre located near the surface of a bulk diamond and to place the sample on the AFM tip \cite{Pelliccione2014,Rugar2015,Haberle2015}. Of these various strategies, attachment of the nanodiamond to the AFM tip is the only one which can provide the detailed topography of the sample in addition to quantum sensor information. This is a key requirement for a practical, versatile instrument capable of imaging complex samples such as assemblies of nanoparticles, nanostructured thin films or biological cells. Furthermore, most probe microscopy experiments to date have employed a single NV centre as the scanning probe, rather than an ensemble. This is motivated by the fact that the ultimate spatial resolution is in principle given by the size of the probe volume, which is minimal for a single NV centre ($\approx 1$ nm), although in practice the resolution is limited by the probe-sample distance, which is typically several tens of nm at least \cite{Rondin2012,Maletinsky2012}. With an ensemble of NV centres, the spatial resolution is degraded by spatial averaging over the ensemble. However, the acquisition time is also dictated by the number $N$ of NV centres employed in the sensor, since this limits the signal-to-noise ratio which scales as $\sqrt{N}$  \cite{Rondin2014}. As a consequence, acquisition times up to several tens of hours per image are typically required with a single NV centre, which imposes serious technical constraints owing to mechanical and thermal drifts present in the microscope \cite{Pelliccione2014,Rugar2015,Haberle2015,Grinolds2013}.

In this work, we seek to optimise the trade-off between acquisition time and resolution by employing a small NV ensemble as our probe. Based on this idea, we demonstrate a scanning nano-spin ensemble microscope that is capable of providing simultaneous topographic and quantum sensor information with a spatial resolution better than 100 nm and typical acquisition times of an hour. We illustrate the versatility of our microscope by applying various sensing modes to two different samples. We first use spin resonance spectroscopy and spin relaxometry to image the magnetic properties of aggregates of maghemite nanoparticles (Fe$_2$O$_3$ in the cubic $\gamma$ phase). Secondly, we demonstrate nanoscale thermal imaging of a photo-heated gold nanoparticle, through spin resonance spectroscopy of the quantum probe in a fluid environment. These results demonstrate that our approach, which relies on nanodiamonds containing a small ensemble of NV spins, is a promising avenue to realise a practical, versatile scanning quantum probe microscope.

The experimental set-up is depicted in Fig. \ref{Fig1}a. A nanodiamond hosting multiple NV centres is attached to the tip of an AFM. A microscope objective lens placed below the sample, which sits on a thin transparent substrate, enables us to excite the NV centres and collect their fluorescence. In addition, a microwave antenna is fabricated on top of the substrate to allow the NV spin resonances to be driven. The quantum probes are type-Ib nanodiamonds $\approx100$ nm in diameter, containing $N\approx100$ NV centres per particle. This leads to a collected fluorescence signal of up to $20\times10^6$ photons per second (see SI) and a gain in signal-to-noise ratio of $\sqrt{N}\approx10$ over a single NV centre probe. By monitoring the fluorescence intensity while sweeping the microwave frequency, a spin resonance spectrum of the NV ensemble is obtained (Fig. \ref{Fig1}b). It generally exhibits two lines at frequencies $\nu_-$ and $\nu_+$. By virtue of Zeeman, Stark and thermal expansion effects, $\nu_-$ and $\nu_+$ are well characterised functions of the local magnetic field ${\bf B}$ \cite{Rondin2014,Gopi2008}, electric field ${\bf E}$ \cite{Dolde2011} and temperature $T$ \cite{Acosta2010}. While spin resonance spectroscopy is sensitive only to the static component -- or sub-kHz variations -- of ${\bf B}$, ${\bf E}$ and $T$, methods have been developed to probe fast fluctuations of these quantities \cite{Maze2008,Hall2009,Cole2009,Meriles2010,deLange2011}. These methods rely on quantum techniques to control and measure the spin dynamics of the NV centres, which are strongly dependent on fluctuations in the local environment. Fig. \ref{Fig1}c illustrates two examples of such measurements, performed on a nanodiamond attached to an AFM tip, termed spin relaxation and spin decoherence measurements, respectively. The spin relaxation time $T_1$ (left-hand graph in Fig. \ref{Fig1}c) is measured by applying laser pulses to initialise and subsequently read out the NV spin state after a wait time $\tau$. The relaxation rate $\Gamma_1=1/T_1$ can be written as the sum $\Gamma_{1}=\Gamma_{1,\rm int}+\Gamma_{1,\rm ext}$ where $\Gamma_{1,\rm int}$ is the intrinsic relaxation rate and $\Gamma_{1,\rm ext}$ is the relaxation caused by the fluctuations of the external magnetic field occurring at the spin resonance frequencies $\nu_\pm\approx3$ GHz \cite{Cole2009,Steinert2013,Tetienne2013,Kaufmann2013,Ermakova2013,Sushkov2014}. The spin coherence time $T_2$ (right-hand graph in Fig. \ref{Fig1}c) is measured by applying MW pulses according to a spin echo sequence. It is sensitive to slower magnetic field fluctuations, in the kHz-MHz range \cite{Rondin2014,McGuinness2013,deLange2011}. In the following, we will use spin resonance spectroscopy to image static magnetic and temperature fields, but also spin relaxometry to image fast magnetic fluctuations. 

Owing to the gain in the collected fluorescence signal of over two orders of magnitude compared to the case for a single NV centre probe, it is possible to conduct imaging experiments with greatly improved acquisition speed. Of particular interest is the use of NV quantum sensing to characterise the magnetic properties of individual magnetic nano-objects. As a demonstration experiment, we image and investigate the dynamics of small aggregates of maghemite ($\gamma$-Fe$_2$O$_3$) nanoparticles. Magnetic nanoparticles have been exploited in a broad range of applications including magnetic seals and inks, magnetic recording media, catalysts, and ferrofluids, as well as in contrast agents for magnetic resonance imaging and therapeutic agents for cancer treatment \cite{Teja2009}. Maghemite nanoparticles in particular exhibit a rich variety of tunable magnetic phenomena, ranging from superparamagnetism below a critical diameter ($\approx10$ nm at room temperature) to ferromagnetism for bigger particles and even superferromagnetism for some ordered assemblies of particles \cite{Teja2009}. However, it remains a challenge to study the magnetic properties of individual -- or small clusters of -- such nanoparticles, especially in their superparamagnetic phase \cite{Schreiber2008,Neves2010}.  

\begin{figure*}[tb!]
\begin{center}
\includegraphics[width=1\textwidth]{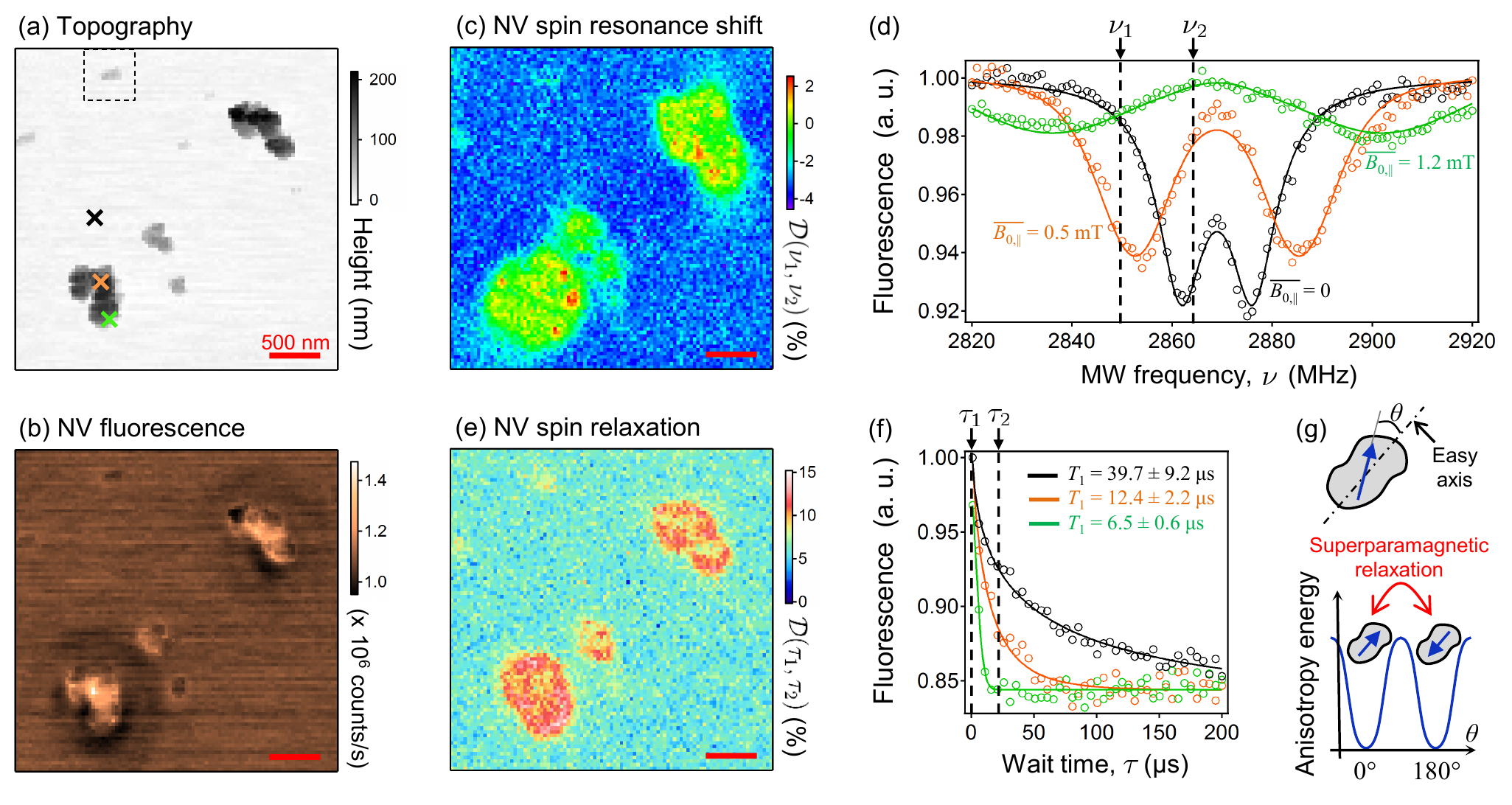}
\caption{Magnetic imaging of small aggregates of maghemite nanoparticles. (a) AFM image of aggregates of maghemite ($\gamma$-Fe$_2$O$_3$) nanoparticles on a glass cover slip. (b) NV fluorescence image. (c) NV spin resonance shift image, corresponding to the differential signal ${\cal D}(\nu_1,\nu_2)$ (see definition in text). The frequencies $\nu_1=2850$ MHz and $\nu_2=2865$ MHz are chosen to give a negative (blue) signal in the absence of magnetic field, and a vanishing or positive (red) signal in the presence of a sizeable magnetic field, as illustrated in (d). (d) Optically detected spin resonance spectra recorded at three different locations as indicated by the crosses in (a) with corresponding colors. The solid lines are the fit to the model described in the text. (e) NV spin relaxation image, corresponding to the differential signal ${\cal D}(\tau_1,\tau_2)$ (see definition in text). The wait times are $\tau_1=1~\mu$s and $\tau_2=20~\mu$s, as illustrated in (f). (f) Spin relaxation curves recorded at the same locations as in (d). The solid lines are the fit to a stretched exponential function, from which we infer the spin relaxation time $T_1$. (g) Schematic illustration of superparamagnetism. The magnetic anisotropy energy of a particle is plotted as a function of the angle $\theta$ between the magnetic moment and the easy (anisotropy) axis of magnetisation. Thermal fluctuations can cause fast reversals of the magnetic moment by overcoming the energy barrier.}
\label{Fig2}
\end{center}
\end{figure*}

Small aggregates of maghemite nanoparticles of diameter $5-25$ nm were prepared on a glass cover slip (see SI). Fig. \ref{Fig2}a shows an AFM image of the sample acquired with a nanodiamond on the tip, revealing aggregates of various sizes ranging from 50 nm to more than 500 nm, corresponding to $10-100$ particles per aggregate. In addition to the topographic image, the fluorescence signal from the NV ensemble is monitored under a given sequence of laser and microwave pulses, and analysed to form the corresponding NV sensor image. For instance, Fig. \ref{Fig2}b shows the fluorescence intensity under continuous laser excitation, without microwave excitation. This provides a near-field optical image of the sample, which correlates well with the topography image. 

The magnetisation of each magnetic particle can be decomposed as ${\bf M}(t)={\bf M}_0+\delta{\bf M}(t)$, where ${\bf M}_0$ is the static component and $\delta{\bf M}(t)$ the fluctuating one. Likewise, the magnetic field produced by ${\bf M}(t)$ at the nanodiamond location is written as ${\bf B}(t)={\bf B}_0+\delta{\bf B}(t)$. To gain information about the magnetic state and dynamics of the particles, we applied two different sensing schemes, which rely on spin resonance spectroscopy to infer ${\bf B}_0$ and spin relaxation to infer $\delta{\bf B}(t)$. In Fig. \ref{Fig2}c, two MW frequencies $\nu_1$ and $\nu_2$ are consecutively applied and the normalised difference, the spin resonance shift, ${\cal D}(\nu_1,\nu_2)=[{\cal F}(\nu_2)-{\cal F}(\nu_1)]/{\cal F}(\nu_2)$ is plotted, with ${\cal F}(\nu_{1,2})$ being the NV fluorescence intensity under MW frequency $\nu_{1,2}$. Due to the Zeeman effect that shifts the spin resonance lines in the presence of a magnetic field, this method provides a map of the static component ${\bf B}_0$ of the magnetic field produced by the sample \cite{Rondin2014}. For a single NV centre, the resonance frequencies are $\nu_{\pm}=D\pm\sqrt{E^2+(\gamma_eB_{0,\parallel}/2\pi)^2}$ where $D$ and $E$ are the zero-field splitting parameters, $\gamma_e/2\pi=28$ GHz/T is the electron gyromagnetic ratio and $B_{0,\parallel}$ is the magnetic field projection along the NV centre's symmetry axis. For an ensemble of NV centres, the spectral features are broadened because the probe experiences four different magnetic field projections at the same time corresponding to the four possible crystallographic axes, and due to variations in the zero-field splitting parameters $D$ and $E$ \cite{Doherty2013}. It is seen in Fig. \ref{Fig2}c that the aggregates produce a measurable static magnetic field component. Fig. \ref{Fig2}d shows spin resonance spectra acquired with the tip positioned at three different locations on the sample, indicated by crosses in Fig. \ref{Fig2}a. The spectra are fitted to a model that assumes a normal distribution of magnetic field amplitudes $B_{0,\parallel}$ with a mean value $\overline{B_{0,\parallel}}$ and standard deviation $\sigma_{B_{0,\parallel}}$ (see SI). The width $\sigma_{B_{0,\parallel}}$ accounts for the averaging over the various orientations and positions of the NV centres contained in the nanodiamond, as well as over the vertical oscillation of the AFM tip. We found these values to reach up to $\overline{B_{0,\parallel}}=1.2$ mT and $\sigma_{B_{0,\parallel}}=0.5$ mT. This indicates the presence of maghemite nanoparticles in a ferromagnetic state, as expected for particles bigger than $\approx10$ nm. 

To probe the dynamics of the magnetisation, we measured the spin relaxation time of the NV ensemble. NV relaxometry has been recently applied to the detection of paramagnetic molecules \cite{Pelliccione2014,Steinert2013,Tetienne2013,Kaufmann2013,Ermakova2013,Sushkov2014}, and superparamagnetic nanoparticles \cite{Schafer2014,Lorch2015}. Fig. \ref{Fig2}e presents a spin relaxation image formed by recording the fluorescence signal ${\cal F}(\tau)$ from the nanodiamond after two different wait times $\tau_1$ and $\tau_2$, and plotting the normalised difference ${\cal D}(\tau_1,\tau_2)=[{\cal F}(\tau_1)-{\cal F}(\tau_2)]/{\cal F}(\tau_1)$. A strong decrease in relaxation time is observed for most aggregates. This is confirmed by the full relaxation curves measured at different locations (Fig. \ref{Fig2}f). The relaxation time is decreased from $T_1=39.7\pm9.2~\mu$s on the bare substrate down to $T_1=6.5\pm0.6~\mu$s for the biggest aggregate in the investigated area. As the $T_1$ relaxation time of the NV centres is governed by the local field fluctuations at the resonance frequencies of $\nu_\pm\approx3$ GHz, our observations suggest that the maghemite aggregates exhibit strong magnetisation fluctuations in the GHz range. Assuming that the fluctuations of $\delta{\bf M}(t)$ can be described by an Ornstein-Uhlenbeck process with a relaxation time $\tau_m$ specific to each particle, the induced relaxation rate is $\Gamma_{1,\rm ext}\approx\frac{\gamma_e^2}{2\pi D}\langle\overline{\delta B(t)^2}\rangle$, where $\langle\overline{\delta B(t)^2}\rangle$ is the variance of the magnetic field amplitude averaged over the NV ensemble produced by the particles satisfying the resonance condition $\tau_m\approx\frac{1}{2\pi D}$ (see SI). The measured NV relaxation time of $T_1\approx6.5~\mu$s therefore translates into a field fluctuation $\langle\overline{\delta B(t)^2}\rangle\approx(270~\mu{\rm T})^2$.

Such fluctuations are consistent with a superparamagnetic behaviour (Fig. \ref{Fig2}g). Indeed, for particles below a certain size, thermal fluctuations can overcome the energy barrier due to magnetic anisotropy, resulting in spontaneous reversals of the magnetisation direction. For uniaxial anisotropy, the superparamagnetic relaxation time is given by $\tau_m=\tau_0\exp\left(\frac{KV}{k_BT}\right)$, where $K$ is the anisotropy constant, $V$ the particle's volume, $k_B$ the Boltzmann constant and $T$ the temperature. For maghemite particles, the prefactor $\tau_0$ lies in the range $10^{-11}-10^{-9}$ s while $K\approx2-7\times10^{4}$ J/m$^3$ \cite{Jonsson1997,Rebbouh2007,Disch2014}. For the smallest particles with a diameter of just 5 nm, this leads to $\tau_m=10^{-11}-10^{-9}$ s at room temperature, which is well within the sensitivity window of NV relaxometry as $\frac{1}{2\pi D}\approx6\times10^{-11}$ s. On the other hand, particles as large as 25 nm in diameter yield $\tau_m\gg1$ s, which correspond to the quasi-static, ferromagnetic state. Our images therefore suggest that most aggregates are comprised of both small superparamagnetic particles and larger ferromagnetic particles. Furthermore, from the relative strength of the static and fluctuating magnetic signals, we deduce that the superparamagnetic particles constitute about 10\% (by volume) of the biggest aggregates in Fig. \ref{Fig2}a (see SI). We note that this experiment closely follows the initial idea in Ref. \cite{Cole2009} to use a scanning NV probe to obtain simultaneous static and dynamic information (through $T_1$) about a sample. Measuring the $T_2$ coherence time would provide further access to low-frequency field fluctuations. However, no significant variation in $T_2$ time was observed in our experiments, which is due to the intrinsic $T_2$ time being much shorter than the intrinsic $T_1$ time \cite{Steinert2013}.

\begin{figure}[h!]
\begin{center}
\includegraphics[width=.45\textwidth]{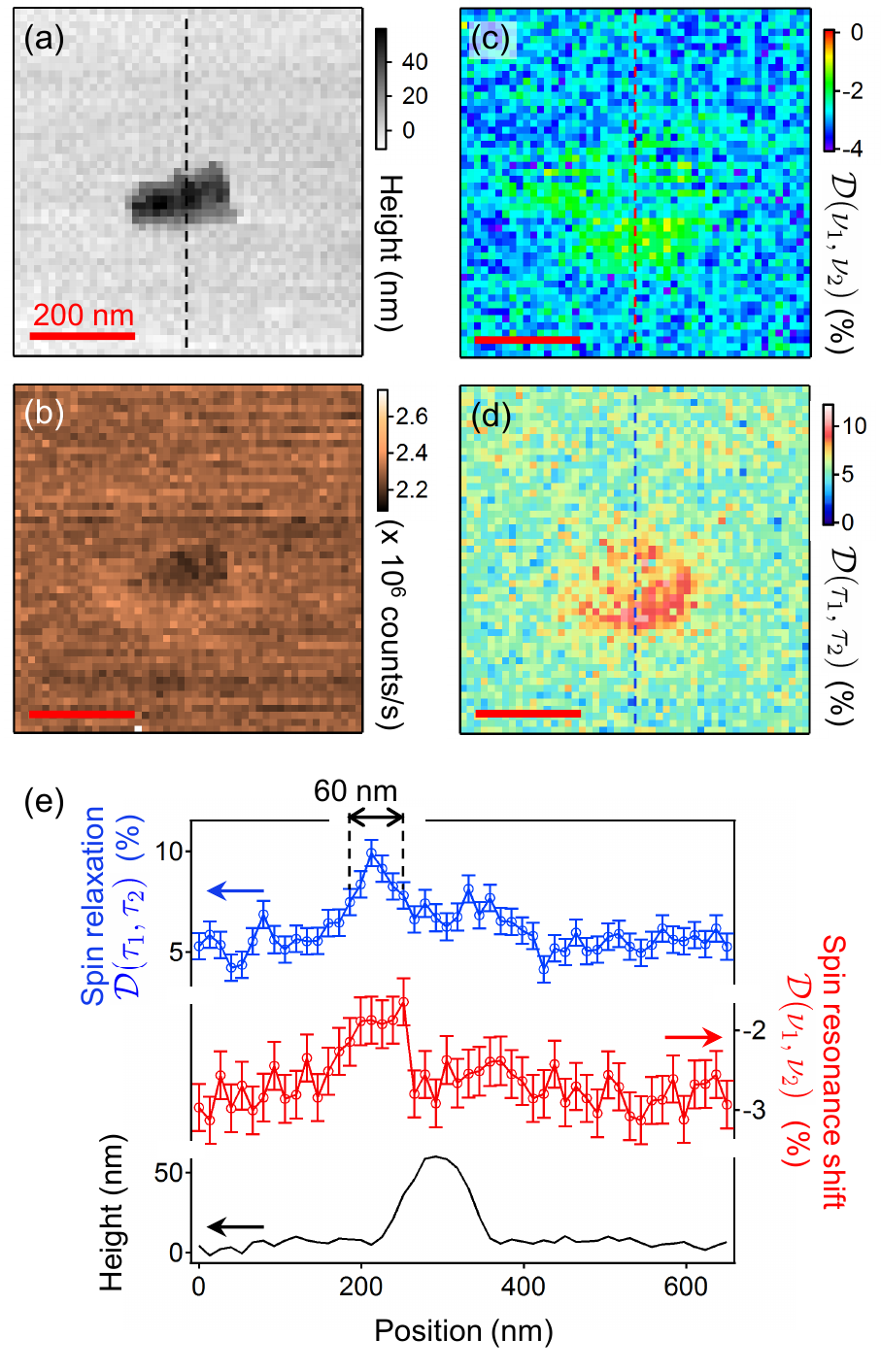}
\caption{Imaging a few-particle maghemite aggregate. (a) AFM image of an aggregate made of a few maghemite nanoparticles. The area corresponds to the dashed square in Fig. \ref{Fig2}a. (b) NV fluorescence image. (c) NV spin resonance shift image. (d) NV spin relaxation image. (c) and (d) are recorded with the same parameters as in Figs. \ref{Fig2}c and \ref{Fig2}e. (e) Line cuts extracted from (a), (c) and (d), taken along the dashed line shown in each corresponding image. The averaging width is 2 pixels.}
\label{Fig3}
\end{center}
\end{figure}

An important advantage of using a multiple-NV sensor is the decreased acquisition time due to increased collected photon signal. The magnetic images shown in Fig. \ref{Fig2}c and \ref{Fig2}e have been acquired with dwell times of 0.4 and 1 s per pixel, respectively, corresponding to a total frame time of 60 and 150 minutes. This allows us to acquire the data in a single pass as in conventional AFM imaging, as opposed to the significantly longer acquisition times typically needed in single NV imaging experiments \cite{Pelliccione2014,Rugar2015,Haberle2015,Grinolds2013}, which require complex multi-pass procedures and re-alignment algorithms. We stress that although our NV ensemble has inferior features compared with a single NV centre in a high-purity bulk diamond (e.g., broadened resonance lines, shorter intrinsic relaxation time) which reduces the sensitivity to {\it small} changes -- i.e., a resonance shift much smaller than the line width or a relative $T_1$ change much smaller than unity --, this does not affect the present experiments where {\it large} changes are observed. Thus, our images fully benefit from the $\sqrt{N}$ increase in signal-to-noise ratio.
Apart from the increased acquisition speed, a further important feature of our microscope is its spatial resolution. By imaging a cluster of a few ($\approx10$) maghemite nanoparticles, whose topography is shown in Fig. \ref{Fig3}a, the spatial resolution of the instrument can be estimated. Figs. \ref{Fig3}b, \ref{Fig3}c and \ref{Fig3}d show the near-field optical, spin resonance shift and spin relaxation images, respectively, obtained with the same nanodiamond probe as in Fig. \ref{Fig2}. The total acquisition time was 8 minutes for Figs. \ref{Fig3}b and \ref{Fig3}c, and 42 minutes for Fig. \ref{Fig3}d. Again, a fast magnetic fluctuation component is observed (Fig. \ref{Fig3}d), while the static field seems comparatively weaker (Fig. \ref{Fig3}c). Detailed analysis of the signal strength confirms that this aggregate is mainly composed of superparamagnetic particles (see SI). A line cut across the aggregate is shown in Fig. \ref{Fig3}e. It reveals a feature in the relaxation data as small as 60 nm (full width at half maximum). This spatial resolution is limited by the effective probe volume, which is given by the size of the nanodiamond ($\approx100$ nm). We note that the magnetic signal is largest on the sides of the aggregate. This is due to the fact that the AFM tip has a small vertical oscillation, therefore the effective probe-sample distance is larger when the tip is above the particles -- hence a smaller mean magnetic field -- than when it is positioned at the side of the particle. The asymmetry is attributed to the non-spherical shape of the nanodiamond.

\begin{figure*}[t!]
\begin{center}
\includegraphics[width=0.9\textwidth]{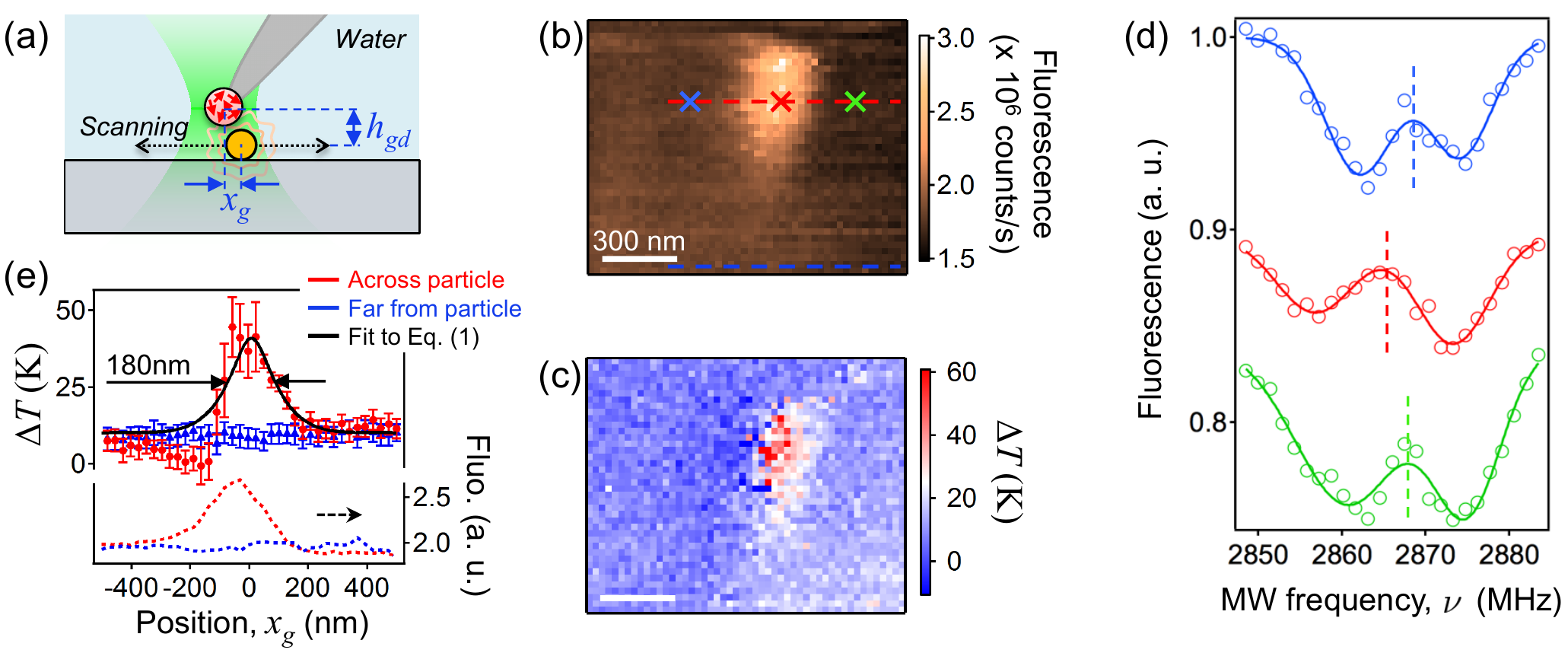}
\caption{Thermal imaging of a photo-heated gold nanoparticle. (a) Schematic of the experiment. To ensure good thermal conductivity between probe and sample, both AFM tip and sample are immersed in water. (b,c) Fluorescence image (b) and temperature map (c) obtained simultaneously by scanning a 40-nm gold particle relative to the nanodiamond probe and its excitation laser. The integration time is 1.5 s per pixel, hence 50 minutes for the whole frame. (d) Optically detected spin resonance spectra corresponding to three different pixels of the scan, located as indicated by the crosses in (b) with matching colours. The solid lines are the fit to a sum of two Lorentzian functions. The vertical dashed lines indicate the splitting parameter $D$ deduced from the fit, from which we infer the temperature increase $\Delta T$ (see text). The spectra are offset for clarity. (e) Line cuts extracted from (b) and (c) taken along the dashed lines shown in (b) with matching colours. The averaging width is 4 pixels. The black solid line is a fit to Eq. (\ref{eqTd}) with the mean diamond-gold particles distance $h_{gd}$ and maximum gold particle temperature $\Delta T_g^{\rm max}$ as free parameters (see text).}
\label{Fig4}
\end{center}
\end{figure*}

Quantum sensing based on NV centres in diamond is not limited to the measurement of magnetic fields. It has recently been shown that the NV spin can be used as a sensitive thermometer by relying on the temperature dependence of the crystal field splitting parameter $D$ \cite{Acosta2010,Toyli2013,Kucsko2013,Neumann2013,Tzeng2015}. Here we extend this work and demonstrate the concept of nanoscale temperature mapping. As a test sample, we use gold nanoparticles of diameter $30\pm10$ nm, which exhibit a surface plasmon resonance at around 530 nm  (see SI). The particles can therefore be efficiently heated using the same laser as that used to excite the NV centres. Under these conditions, the gold nanoparticles act as nanoscale heat sources. A sample was prepared by spin coating gold nanoparticles onto a glass cover slip (see SI). Because the glass substrate has a thermal conductivity much greater than that of air, any heat generated by illumination of the gold particle in air would be expected to spread preferentially towards the substrate, resulting in a negligible change in the air temperature above the particle. Hence, in order to probe this photo-induced heating, the nanodiamond probe and the sample were immersed in water, as schematically shown in Fig. \ref{Fig4}a. Under these conditions, the heat spreads quasi-isotropically around the particle, enabling the nanodiamond sensor to report the local temperature change. 

In order to locate a gold particle and scan the region around it with the nanodiamond probe tip, we relied on the collected fluorescence rather than on the topography. The gold particles produce a weak but measurable fluorescence signal, which is maximised when the nanodiamond is directly above the particle as the laser beam is always focused on the nanodiamond (see Fig. \ref{Fig4}a). Fig. \ref{Fig4}b shows a fluorescence image obtained by scanning the nanodiamond around a 40-nm-diameter gold particle (see SI). The fluorescent spot corresponds to the gold particle and indicates a significant absorption of the laser light, which is expected to turn primarily into heat. To form a temperature map, a full spin resonance spectrum is recorded at each pixel of the scan. The spectrum is acquired under continuous laser excitation, with an incident power of 250 $\mu$W, in order to maximise the heating. From the fits to the spectroscopy data, we then infer the splitting parameter $D$ and deduce the temperature increase $\Delta T=T-T_0$, where $T_0$ is the temperature measured in the absence of laser heating (see SI). The resulting temperature map is shown in Fig. \ref{Fig4}c. It reveals a region of increased temperature up to $\approx40$ K, which is spatially correlated with the fluorescence spot (Fig. \ref{Fig4}b). Individual spectra corresponding to a few selected pixels are shown in Fig. \ref{Fig4}d. A negative shift of the resonance frequencies is clearly observed at the hot spot. A line cut across the particle extracted from the temperature map is shown in Fig. \ref{Fig4}e (red data), indicating a peak of full width at half maximum $\approx180$ nm. Also shown in Fig. \ref{Fig4}e is a line cut taken far from the particle (blue data). A small amount of background heating of $\Delta T_{\rm bkg}\approx9$ K is evident, suggesting that the nanodiamond is heated by the laser even in the absence of the gold particle, most likely through heating of the silicon AFM tip. We stress that the increase in fluorescence signal caused by the use of an NV ensemble is crucial to this particular experiment. Indeed, the signal from a single NV centre would be overwhelmed by the background fluorescence produced by the gold particle, making the measurement highly challenging using the conventional single NV approach.

To gain more insight into the heating of the gold particle, we derived an analytical expression for the nanodiamond temperature $\Delta T_d$ as a function of the relative position $x_g$ of the gold particle (see SI). The temperature increase of the gold particle is denoted $\Delta T_g$ and is assumed to be proportional to the laser intensity at position $x_g$, while the temperature distribution outside the gold particle is obtained from the thermal diffusion equation in the steady-state regime, assuming for simplicity a spherical particle in a homogeneous medium \cite{Baffou2010}. This yields  
\begin{equation} \label{eqTd}
\Delta T_d(x_g)\approx\Delta T_{\rm bkg}+\Delta T_g^{\rm max}\exp\left(-\frac{x_g^2}{2\sigma_L^2}\right)\frac{R_g}{\sqrt{x_g^2+h_{gd}^2}},
\end{equation} 
where $\Delta T_g^{\rm max}$ is the temperature of the gold particle when it is in the centre of the laser beam (i.e., $x_g=0$), $\sigma_L\sqrt{8\ln2}=300$ nm is the laser beam width, $R_g=20$ nm is the gold particle radius, and $h_{gd}$ is the centre-to-centre distance between the gold particle and the nanodiamond probe at $x_g=0$ (see Fig. \ref{Fig4}a). Fitting Eq. (\ref{eqTd}) to the data gives $h_{gd}=70\pm20$ nm, consistent with the known sizes of the gold and diamond particles (radius of 20 and 50 nm, respectively), and $\Delta T_g^{\rm max}=110\pm30$ K (Fig. \ref{Fig4}e). This temperature increase can be compared to the theoretical value $\Delta T_g^{\rm max}=P_{\rm abs}/(4\pi R_g\kappa)$ where $P_{\rm abs}$ is the absorbed power, which applies to the case of an isolated particle in a homogeneous medium of thermal conductivity $\kappa$. Using the bulk thermal conductivity of glass $\kappa=0.8$ W$\cdot$m$^{-1}\cdot$K$^{-1}$ and a power $P_{\rm abs}=9~\mu$W calculated from Mie scattering theory \cite{BohrenHuffman}, one predicts $\Delta T_g^{\rm max}=40$ K. This is smaller than the value of 110 K inferred from the experiment, suggesting there may be enhanced absorption by the gold nanoparticle in the experiment due to, e.g., a non-spherical particle shape or tip-induced field concentration effects. Nevertheless, these data illustrate that our technique can be used to investigate and measure nanoscale heating and thermal gradients induced by light, chemical reactions or other energy dissipation sources. In addition, Eq. (\ref{eqTd}) shows that the linewidth of 180 nm is a convolution of both the temperature profile above the gold particle at a given position of the laser beam and the profile of the stimulating laser beam itself. This width is therefore related to the measurement conditions rather than to the intrinsic spatial resolution, which is limited only by the probe volume of $\approx 100$ nm corresponding to the nanodiamond size. We note that direct mapping of the temperature profile during constant heating could be achieved by employing an additional heating laser, thereby providing a way to investigate photo-induced heating with a spatial resolution drastically better than the current, diffraction-limited techniques \cite{Baffou2012,Donner2012}. 

In this work we have demonstrated a new approach to scanning quantum probe microscopy that relies on a spin ensemble in a nanodiamond. In addition to conventional AFM topography, our microscope is capable of providing spin resonance and spin relaxation images with typical acquisition times of an hour. Our instrument is capable of imaging both the static and fluctuating magnetic fields generated by small aggregates of maghemite nanoparticles, enabling detection of both ferromagnetic and superparamagnetic particles. The nano-spin ensemble is also shown to be a sensitive nanoscale thermometer providing detailed temperature maps of photo-heated gold particles. 

The present approach and experimental apparatus is particularly suited to the study of biological samples for several reasons. First, scanning nanodiamond microscopy is compatible with operation in water, which paves the way to imaging experiments on living samples in their native environment. Second, the use of multiple NV centres provides a significant gain in acquisition speed, which is a clear advantage for fragile samples that cannot tolerate long acquisition times. In the experiments reported here the nanodiamond contained $\approx100$ NV centres, but even higher concentration of NV centres could be used to further improve the scanning speed \cite{Chang2008}. Lastly, our instrument provides topographic images correlated to the quantum sensor images, with a combined spatial lateral resolution below 100 nm. This is a crucial requirement for experiments performed on samples with a non-trivial shape. We envisage using this technology to investigate spin signals in cell membranes stemming, e.g., from spin labels \cite{Kaufmann2013} or ion channels \cite{Hall2010}, which could give crucial insight into cell membrane functions and dynamics. 

On the other hand, scanning quantum probe microscopy also opens up opportunities for the study of solid-state samples. In particular, the temperature mapping capability reported in this paper provides a new platform for studying thermal transfer at the nanoscale. This is particularly relevant in the race towards ever smaller micro/nano-electronic devices, where the precise characterisation of heat transfer remains a fundamental bottleneck \cite{Mecklenburg2015}. 

\subsection*{Acknowledgements}

The authors thank V. Jacques, A. Stacey and L. H. Hall for fruitful discussions, B. Moubaraki for performing the SQUID measurements, and H. C. Chang for the supply of nanodiamonds. This work was supported in part by the Australian Research Council (ARC) under the Centre of Excellence scheme (project No. CE110001027) and through a LIEF grant (project No. LE110100161). LCLH acknowledges the support of an ARC Laureate Fellowship (project No.  FL130100119). 

\begin{widetext}

\vspace{0.5cm}

\begin{center}
{\large SUPPORTING INFORMATION}
\end{center}

\section{Experimental setup}

The experimental apparatus is based on an AFM (Asylum Research MFP-3D-BIO) mounted on a customised inverted optical microscope (Olympus IX-71). The microscope objective lens (Olympus PlanApoN 60x, NA = 1.42) is mounted on a 3-axis $XYZ$ scanning stage (PI P545-3R7) to allow fast laser scanning. The filter cube (Olympus BX2) loaded in the microscope turret is equipped with a dichroic beam splitter (Chroma ZT532rdc) and an emission filter (Chroma NC359095 ET690/120x). The excitation laser (Laser Quantum Gem 532) is modulated using an acousto-optic modulator (AA Opto-Electronic MT-200-VIS) in a double pass configuration. The modulated laser beam feeds the back port of the Olympus IX-71 microscope through a single-mode fibre. The filtered fluorescence light exits the microscope through the side port and is coupled into a multi-mode fibre connected to a single photon    counting module (Excelitas SPCM-AQRH-14-FC). The photon signal is analysed by multi-function devices (NI-DAQ PCIe-6323 and PCIe-6321) for simple counting and a time digitizer (FastComTec MCS6A-2T4) for time-correlated measurements. 

The sample is mounted onto a 2-axis $XY$ scanning stage while the AFM cantilever sits on a $Z$ piezo-electric actuator for distance control. The NI-DAQ devices are also used to generate analog voltages that control the $XYZ$ objective stage and the $XY$ AFM stage, as well as to read out the $Z$ position of the AFM cantilever, which is directly regulated by the AFM controller. The sample is equipped with a microwave antenna fabricated by optical lithography and metal deposition. It is connected to the output of a microwave generator (Rohde \& Schwarz SMB100A) after modulation (Mini-Circuits ZASWA-2-50DR+) and amplification (Mini-Circuits ZHL-16W-43+). The laser and microwave modulations are controlled by a programmable pulse generator (SpinCore PulseBlasterESR-PRO 500 MHz). All instruments and devices are controlled with a single purpose-built Labview program. 

\section{Probe preparation and characterisation}

\begin{figure*}[t]
\begin{center}
\includegraphics[width=0.6\textwidth]{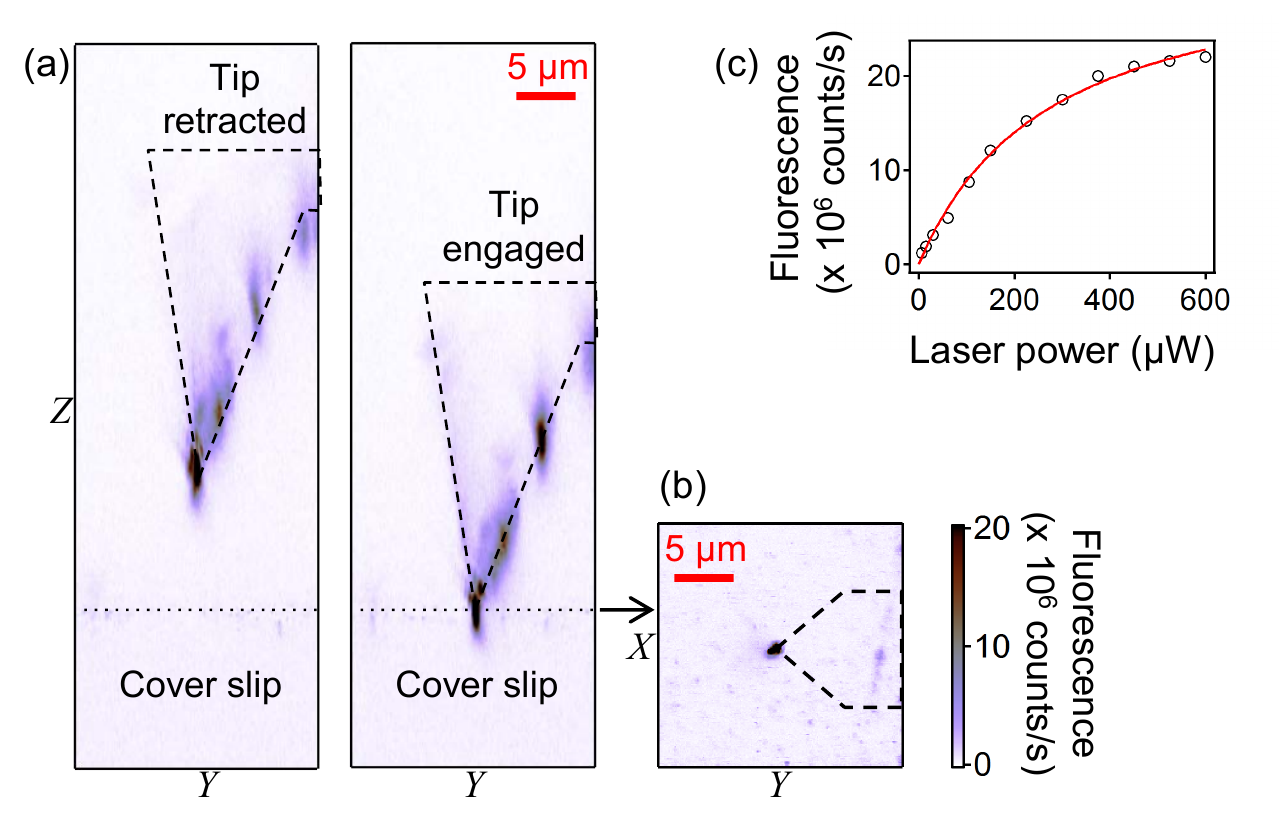}
\caption{(a,b) Confocal images of the AFM tip in the vertical $YZ$ plane (a) and horizontal $XY$ plane (b), the $XYZ$ reference frame being defined in main-text Fig. 1. The dashed (dotted) lines depict the AFM tip (the sample's top surface). The bright spot at the apex of the tip corresponds to the grafted nanodiamond. In (a), the left-hand (right-hand) image shows the tip retracted (engaged) relative to the sample's surface. (c) Number of photons collected per second ${\cal F}(P_L)$ from a nanodiamond on tip as a function of the incident laser power $P_L$. The solid line is the fit to a saturation law ${\cal F}(P_L)=\frac{{\cal F}_\infty}{1+P_{\rm sat}/P_L}$, yielding the fit parameters ${\cal F}_\infty=33.3\pm1.4\times10^6$ counts/s and $P_{\rm sat}=275\pm26~\mu$W.}
\label{FigS1}
\end{center}
\end{figure*} 

The nanodiamonds used in this work (rFND-100) were provided by the BioDiamond team of Academia Sinica (Taipei, Taiwan). They were spin cast on a glass coverslip and characterised both optically and by AFM. The nanodiamonds selected to be attached to the AFM tip were $\approx100$~nm in size and contained between 20 and 100 NV centres based on the observed fluorescence intensity. The AFM probes used here are either medium-soft silicon cantilevers with a nominal tip radius of 9 nm (Olympus AC240TS-R3) for the magnetic imaging experiments or stiff silicon cantilevers with a 2-$\mu$m-diameter plateau tip (Nanosensors PL2-NCHR-10) for the thermal imaging experiments.

To transfer a nanodiamond to an AFM tip, we must first align the laser beam onto the AFM tip. A coarse alignment is achieved using a CCD camera placed on top of the AFM head, providing a precision of $\pm2~\mu$m. The alignment is then fine tuned by correlating the AFM and confocal images obtained when scanning the sample. Once a nanodiamond suitable for sensing -- i.e. with a good spin contrast -- is found, an AFM scan centred on this nanodiamond is performed repeatedly. By decreasing the oscillation amplitude of the tip in AC (tapping) mode or by switching to contact mode, the effective interaction strength between the tip and the nanodiamond is increased, enabling the tip to push the nanodiamond on the substrate and eventually attach it \cite{Rondin2012}. The nanodiamond grafting is evidenced by a strong increase in the fluorescence signal from the tip -- the laser being aligned on the tip at all times -- and an increase in the tip height. To test and strengthen the bonding, we then perform long, low-speed scans in AC mode.

Figs. \ref{FigS1}a and \ref{FigS1}b show confocal images of an AFM tip after grafting a nanodiamond onto it. The nanodiamond appears as a bright fluorescence spot at the apex of the tip. Fig. \ref{FigS1}c shows a typical saturation curve from this spot, that is, the collected fluorescence signal as a function of the incident laser power. In this example the signal reaches up to $22\times10^6$ counts/s close to saturation, with a laser power of 600 $\mu$W. Once a nanodiamond is attached to the AFM tip, its performances as a quantum sensor can be characterised. Fig. \ref{FigS2} shows the relevant characteristic curves for the nanodiamond probes used in this work, the top row corresponding to the probe `ND1' used in main-text Figs. 2 and 3 for the magnetic imaging experiments (Fig. \ref{FigS2}a), while the bottom row corresponds to the probe `ND2' used in main-text Fig. 4 for the thermal imaging experiments (Fig. \ref{FigS2}b). The left-hand column in Fig. \ref{FigS2} corresponds to the spin resonance spectra ${\cal F}(\nu)$, i.e. the fluorescence intensity as a function of the MW frequency. They are fitted to a sum of two Lorentzian lines
\begin{equation} \label{Eq1}
{\cal F}(\nu)={\cal F}_0\left[1-\frac{C_-}{1+\left[\frac{2(\nu-\nu_-)}{\Delta\nu_-}\right]^2}-\frac{C_+}{1+\left[\frac{2(\nu-\nu_+)}{\Delta\nu_+}\right]^2}\right],
\end{equation}
where ${\cal F}_0$ is the off-resonance fluorescence intensity, and $\nu_\pm$, $C_\pm$ and $\Delta\nu_\pm$ are the central frequency, relative contrast and full width at half maximum, respectively, of each of the two spin resonances denoted by the subscript $\pm$. In the absence of external magnetic field, the resonance frequencies are $\nu_\pm=D\pm E$, $D$ and $E$ being the so-called zero-field splitting parameters of the NV centre. Note that since the nanodiamonds contain an ensemble of NV centres, the spectra indicate the mean characteristics of this ensemble. Here we find $D=2868.9\pm0.1$ MHz and $E=7.5\pm0.1$ MHz for ND1 and $D=2868.0\pm0.2$ MHz and $E=5.3\pm0.2$ MHz for ND2. Under typical experimental conditions, the contrast $C_\pm$ is about 8\% and the linewidth is $\Delta\nu_\pm\approx10$ MHz. 

\begin{figure*}[t]
\begin{center}
\includegraphics[width=0.8\textwidth]{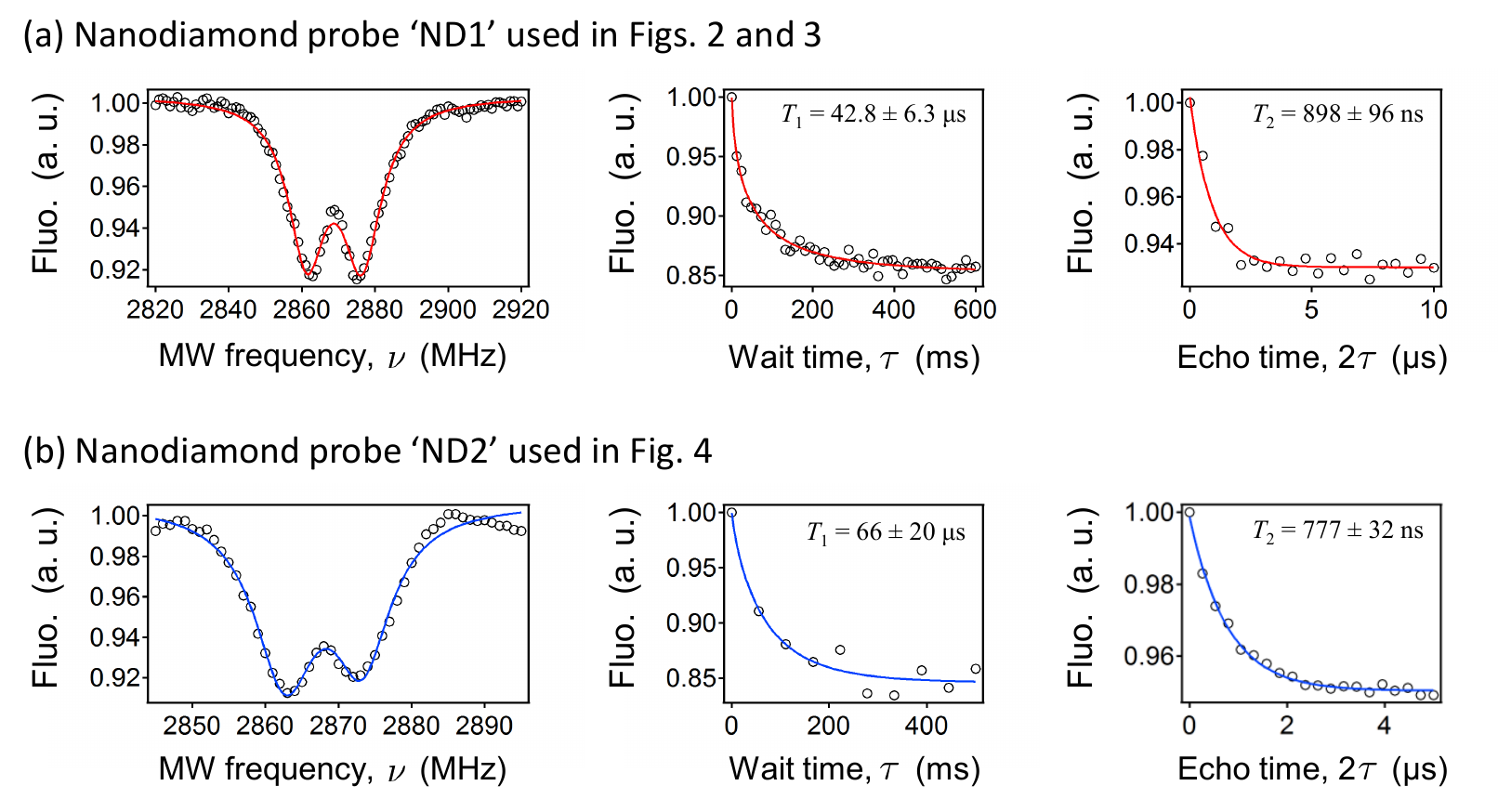}
\caption{Characteristic curves measured for ND1 (a) and ND2 (b). The graphs in the left, middle and right columns correspond to optically detected spin resonance spectra, spin relaxation curves and spin echo decoherence curves, respectively. The markers are the data point while the solid lines are fits as described in the text.}
\label{FigS2}
\end{center}
\end{figure*}

The middle column in Fig. \ref{FigS2} is the spin relaxation curve ${\cal F}(\tau)$ obtained by measuring the fluorescence intensity following an initialisation laser pulse and a wait time $\tau$. The resulting curves are fitted to a stretched exponential function
\begin{equation}
{\cal F}(\tau)={\cal F}_0\left[1-C+C{\rm e}^{-\left(\frac{\tau}{T_1}\right)^n}\right],
\end{equation} 
where ${\cal F}_0$ is the fluorescence intensity in the initialised spin state, $C$ is the relative contrast, $T_1$ is the spin relaxation time of the NV ensemble and $n$ is an exponent that accounts for the inhomogeneity of the spin ensemble. In our nanodiamonds we have typically $C\approx15\%$, $n\approx0.5-1$ and $T_1\approx40-400~\mu$s. The $T_1$ values of ND1 and ND2 are indicated in Fig. \ref{FigS2}.

Finally, The right-hand column in Fig. \ref{FigS2} is the decoherence curve obtained under a spin echo sequence $\frac{\pi}{2}-\tau-\pi-\tau-\frac{\pi}{2}$. The resulting curves are fitted to a single exponential function ${\cal F}(2\tau)={\cal F}_0[1-C+C\exp(-2\tau/T_2)]$, with $T_2$ being the spin coherence time. In our nanodiamonds, $T_2$ is typically comprised between 500 ns and 1.5 $\mu$s in zero external magnetic field.

\section{Supporting information for the magnetic imaging experiments}

\subsection{Sample preparation and characterisation}

The magnetic nanoparticles used for the experiments reported in main-text Figs. 2 and 3 were purchased from Sigma-Aldrich (reference 544884). They are received in the form of a nanopowder of Fe$_2$O$_3$ primarily in the $\gamma$ crystalline phase, or so-called maghemite. We diluted the powder in Milli-Q water to obtain a solution of concentration of 0.1 g/L. After plasma cleaning a glass cover slip for 10 minutes, $50~\mu$L of the maghemite solution is dropped on the cover slip and immediately spinned using a spin caoter with an acceleration of 200 rpm/s and a speed of 3000 rpm for 60 s. These parameters result in a dispersion of particles on the substrate that are mostly isolated from each other, with only little aggregation, as shown in the AFM image in Fig. \ref{FigS3}a. The height of a particle in the AFM image gives an estimate of its mean diameter (Figs. \ref{FigS3}b and \ref{FigS3}c). We find that the diameter of the particles in this sample ranges from 3 nm to 25 nm. This is in good agreement with the surface area of $50-245$ m$^2$/g specified by the manufacturer, which corresponds to a diameter of $5-25$ nm. To obtain aggregates of the particles, we repeat the above procedure but with a spinning speed decreased to 1500 rpm, thus increasing the probability for aggregation. An AFM image of the sample after this step is shown in main-text Fig. 2a.


\begin{figure*}[b]
\begin{center}
\includegraphics[width=0.8\textwidth]{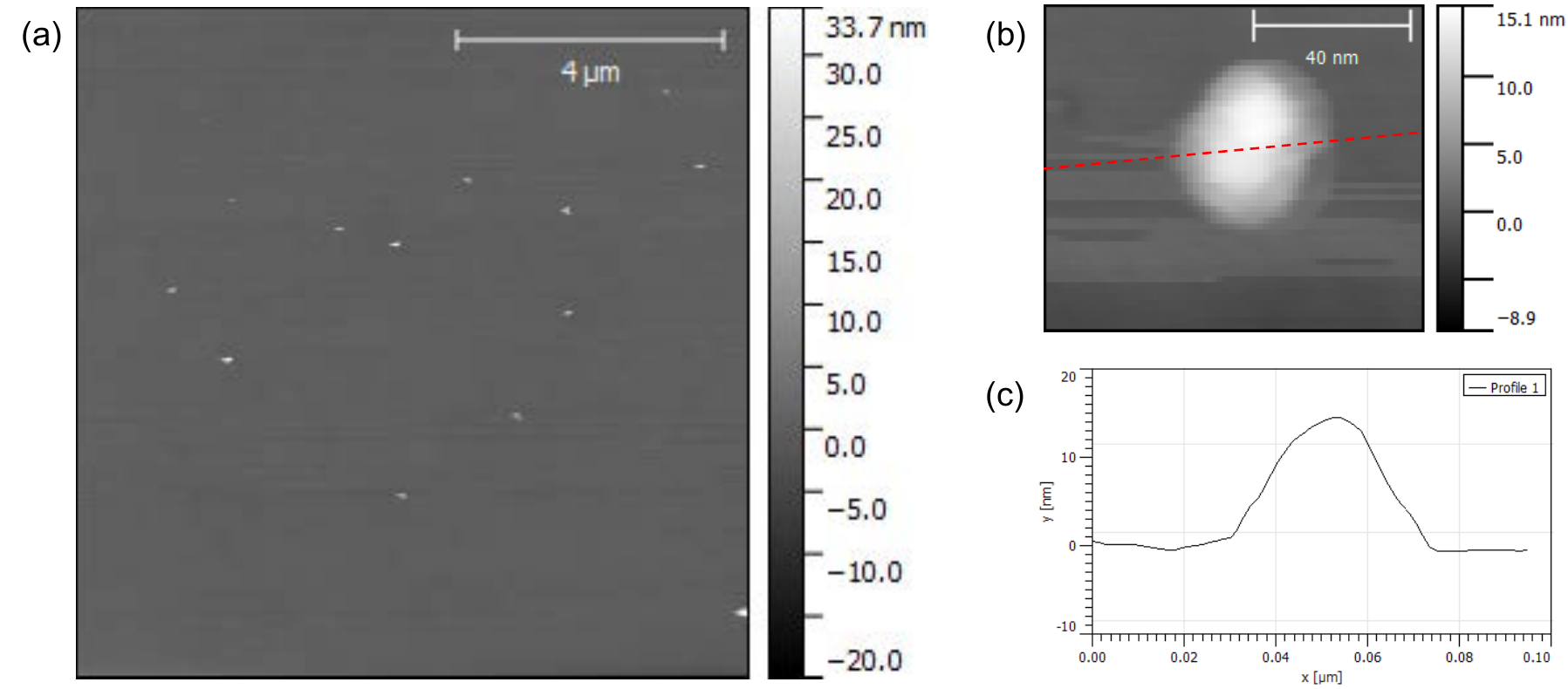}
\caption{(a) AFM image of a sample of non-aggregated maghemite nanoparticles. (b) Close-up view of a single particle. (c) Line cut taken along the red dashed line in (b).}
\label{FigS3}
\end{center}
\end{figure*}

We also characterised the magnetic properties of a macroscopic sample of the same maghemite nanoparticles. We placed $\approx1.6$ mg of the as-received powder in a SQUID magnetometer and measured the average magnetisation $M$ as a function of magnetic field $H$ and temperature $T$. Fig. \ref{FigS3b}a shows the $M(T)$ curve under an applied field of $\mu_0H=1$ T, which should saturate all particles regardless of their size. The small decrease of $M$ at 300 K relative to the low temperature value indicates that the Curie temperature is well above 300 K. Fig. \ref{FigS3b}b shows the $M(H)$ curve at $T=300$ K, obtained by ramping down the field from 5 T to 0. The remnant magnetisation at $H=0$ indicates the presence of ferromagnetic-like particles. For an ensemble of ferromagnetic particles with randomly oriented easy axes, one would expect a ratio $\frac{M(H=0)}{M(H\rightarrow\infty)}\approx\frac{1}{2}$. On the contrary, if the particles were purely superparamagnetic, the magnetisation would average to zero over the measurement time, leading to $\frac{M(H=0)}{M(H\rightarrow\infty)}\approx0$. The present data (Fig. \ref{FigS3b}b) suggests that the powder comprises both ferromagnetic-like (in fact, ferrimagnetic) and superparamagnetic particles, with approximately equal volumes since $\frac{M(H=0)}{M(H\rightarrow\infty)}\approx\frac{1}{4}$.

\begin{figure*}[t]
\begin{center}
\includegraphics[width=0.8\textwidth]{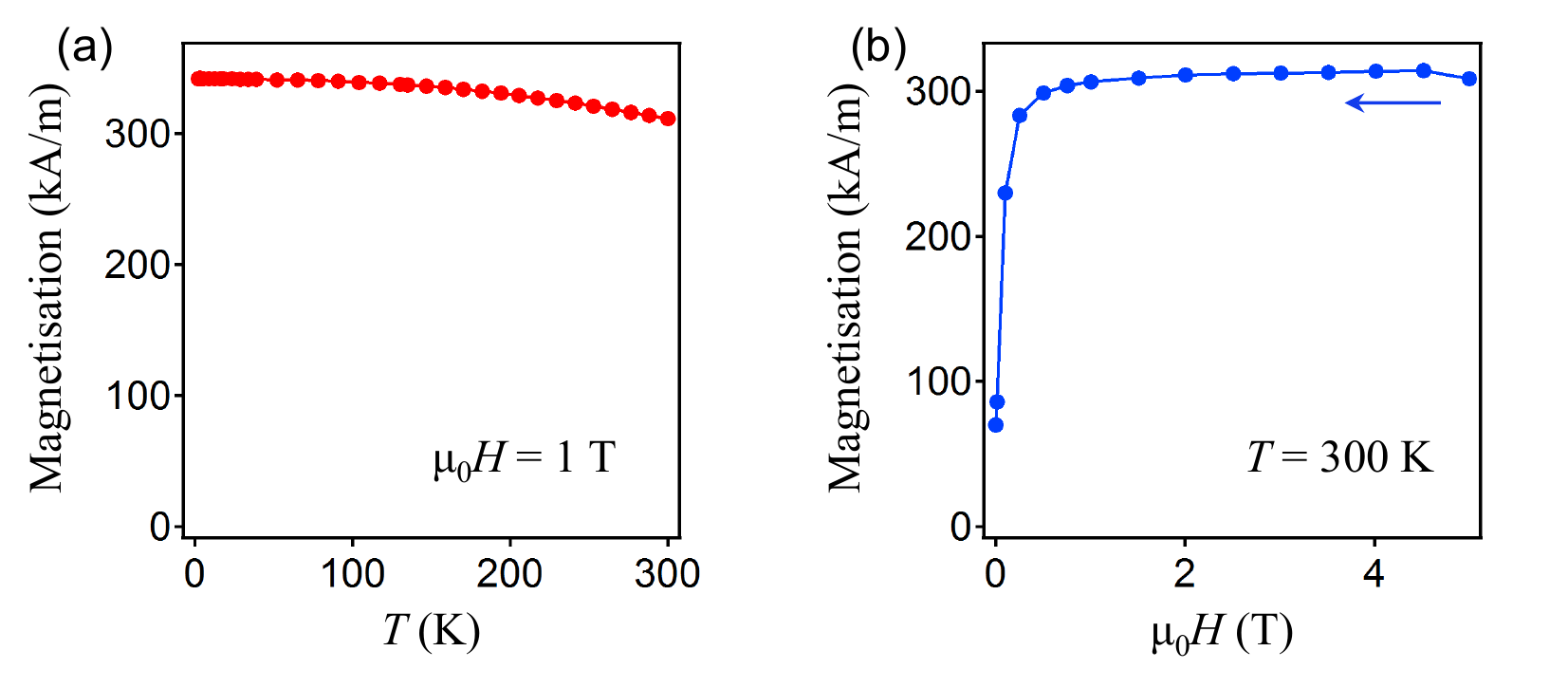}
\caption{(a,b) Characteristic curves of a macroscopic sample (1.6 mg) of the maghemite particles measured using a SQUID magnetometer. (a) Magnetisation versus temperature, under an applied field of 1 T. (b) Magnetisation versus applied field at 300 K. The field is ramped down from 5 T to 0.}
\label{FigS3b}
\end{center}
\end{figure*}

\subsection{Data acquisition methods}

The images reported in main-text Figs. 2 and 3 were obtained by scanning the sample stage while operating the AFM tip in AC (tapping) mode. The amplitude of the vertical oscillation was estimated from force curves to be $\approx 60$ nm in scanning conditions. The NV fluorescence signal is integrated only once for a given time at each pixel, after which the sample is moved to the next pixel. After each line of the scan, the $XYZ$ position of the objective lens is adjusted in order to maximise the NV fluorescence signal. For the spin resonance images (main-text Figs. 2c and 3c), the microwave is switched on for 150 ns, corresponding to a $\pi$ flip rotation of the NV spins, followed by a 300-ns laser pulse for readout and a 1-$\mu$s wait time. This sequence is repeated continuously during the pixel time, with the microwave frequency set to $\nu_1$ for half the pixel time and $\nu_2$ for the second half. The integrated fluorescence signals ${\cal F}(\nu_1)$ and ${\cal F}(\nu_2)$ are recorded using the NI-DAQ device. For the spin relaxation images (main-text Figs. 2e and 3d), a sequence of three 3-$\mu$s-long laser pulses with wait times $\tau_1$ and $\tau_2$ is repeated continuously during the pixel time. The fluorescence signal is acquired using the FastComTec time digitizer and the first 300 ns of each laser pulse are integrated to give the values ${\cal F}(\tau_1)$ and ${\cal F}(\tau_2)$. In these experiments, the incident laser power was $P_L\approx150~\mu$W. All measurements were performed at room temperature under normal atmosphere, with no external magnetic field applied. 

\subsection{Analysis of the spin resonance spectra} \label{secODMR}

Here we describe the model used to analyse the spin resonance spectra recorded on the maghemite aggregates (main-text Fig. 2d). The static component of the magnetic field at the location of NV centre $i$ is denoted ${\bf B}_0^i$. This NV centre will exhibit spin resonances at frequencies
\begin{equation} \label{Eq2}
\nu_{\pm}^i=D\pm\sqrt{E^2+\left(\frac{\gamma_e}{2\pi}B_{0,\parallel}^i\right)^2},
\end{equation}
where $B_{0,\parallel}^i={\bf B}_0^i\cdot\hat{e}_z^i$ is the magnetic field projection along the NV centre's symmetry axis, defined by the unit vector $\hat{e}_z^i$. Experimentally, one probes a ensemble of NV centres with different symmetry axes and positions within the nanodiamond of diameter $\approx100$ nm. In addition, the AFM tip is operated in tapping mode, which implies the existence of a vertical oscillation of $\approx60$ nm at a frequency of $\approx60$ kHz. These effects all contribute to produce a broad distribution of field projections $B_{0,\parallel}^i$, resulting in a broad distribution of frequencies $\nu_{\pm}^i$. We characterise this distribution by a mean value $\overline{B_{0,\parallel}}$ and standard deviation $\sigma_{B_{0,\parallel}}$. That is, the probability of having a field projection $B_{0,\parallel}$ in the NV ensemble during the measurement is given by
\begin{equation} \label{Eq3}
{\cal P}(B_{0,\parallel})=\frac{1}{\sigma_{B_{0,\parallel}}\sqrt{2\pi}}{\rm e}^{-\frac{(B_{0,\parallel}-\overline{B_{0,\parallel}})^2}{2\sigma_{B_{0,\parallel}}^2}}.
\end{equation} 
The spectra in main-text Fig. 2d were fitted by a superposition of spectra in the form of Eq. (\ref{Eq1}) with frequencies given by Eq. (\ref{Eq2}) and with the probability distribution (\ref{Eq3}). The only fit parameters are $\overline{B_{0,\parallel}}$ and $\sigma_{B_{0,\parallel}}$. All other parameters ($D$, $E$, $C_\pm$ and $\Delta\nu_\pm$) are extracted from a reference spectrum recorded far from any magnetic particle (black curve in main-text Fig. 2d).

It is useful to relate the mean field projection $\overline{B_{0,\parallel}}$ to the mean field amplitude $\overline{B_{0}}$. Let us write the field projection seen by a given NV centre as $B_{0,\parallel}^i=B_0\cos\theta_i$, with $\theta_i$ being the angle between the local field ${\bf B}_0$ and the NV centre's axis. Assuming that all directions are equiprobable within the $4\pi$ solid angle, we obtain
\begin{equation}
\overline{B_{0,\parallel}}=\frac{\overline{B_{0}}}{2}.
\end{equation}
The maximum field measured while scanning the sample of maghemite aggregates corresponds to the spectrum shown in main-text Fig. 2d (green curve). The fit to the data yields a mean projection $\overline{B_{0,\parallel}}=1.2\pm0.1$ mT, hence an amplitude $\overline{B_{0}}=2.4\pm0.2$ mT, with a width $\sigma_{B_{0,\parallel}}=0.5\pm0.1$ mT.

\subsection{Analysis of the spin relaxation data} \label{secT1}

We now describe the relation between the measured spin relaxation time ($T_1$) and the local magnetic field fluctuations. The fluctuating component of the magnetic field at the location of NV centre $i$ is denoted $\delta{\bf B}^i(t)$. Its spin relaxation rate reads \cite{Tetienne2013}
\begin{equation}
\Gamma_{\rm tot}^i=\Gamma_{\rm int}^i+\Gamma_{\rm ext}^i
\end{equation}
where
\begin{equation}
\Gamma_{\rm ext}^i=\frac{3\gamma_e^2}{4\pi D}\langle \delta B_\perp^i(t)^2\rangle f(2\pi D\tau_m).
\end{equation}
Here $\langle\ldots\rangle$ designates the time average, $\delta B_\perp^i(t)^2=\delta B_x^i(t)^2+\delta B_y^i(t)^2$ is the transverse component of the field ($z$ being defined by the NV centre's symmetry axis) and $\tau_m$ is the correlation time of $\delta B_\perp^i(t)$. The function $f(x)$ is defined as
\begin{equation}
f(x)=\frac{2x}{1+x^2},
\end{equation}
which is peaked around $x=1$. Since the field $\delta B_\perp^i(t)$ is produced by an assembly of magnetic nanoparticles with different sizes hence different dynamics $\tau_m$, the relaxation rate $\Gamma_{\rm ext}^i$ will be dominated by the particles for which $\tau_m\approx\frac{1}{2\pi D}$, i.e. maximising $f(2\pi D\tau_m)$. One can therefore write
\begin{equation}
\Gamma_{\rm ext}^i\approx\frac{3\gamma_e^2}{4\pi D}\langle \delta B_\perp^i(t)^2\rangle
\end{equation}
where $\delta B_\perp^i(t)$ now accounts only for the contribution of the particles satisfying $\tau_m\approx\frac{1}{2\pi D}$. Averaging over all NV centres and assuming that all directions of the NV axis are equiprobable, we obtain an effective relaxation rate for the nanodiamond
\begin{equation}
\Gamma_{\rm ext}\approx\frac{\gamma_e^2}{2\pi D}\langle \overline{\delta B(t)^2}\rangle,
\end{equation}
where $\delta B(t)^2$ is the squared amplitude of the field seen by a given NV centre and $\overline{\cdots}$ designates the average over the ensemble of NV centres within the nanodiamond.  We can therefore infer this amplitude from the measured relaxation rate according to
\begin{eqnarray}
\langle\overline{\delta B(t)^2}\rangle \approx \frac{2\pi D}{\gamma_e^2}\Gamma_{\rm ext}.
\end{eqnarray}

The maximum relaxation rate measured while scanning the sample of maghemite aggregates corresponds to the data shown in main-text Fig. 2f (green curve). By comparing with the reference curve (black curve), one deduces $\Gamma_{\rm ext}=\Gamma_{\rm tot}-\Gamma_{\rm int}=129\pm15\times10^3$ s$^{-1}$. This translates into a root-mean-square (rms) amplitude $\sqrt{\langle\overline{\delta B(t)^2}\rangle}\approx270~\mu{\rm T}$-rms.

Finally, we can use the ratio between to the static field amplitude $\overline{B_{0}}$ (up to $\approx2.4$ mT in main-text Fig. 2c) and the fluctuating component $\langle\overline{\delta B(t)^2}\rangle$ to estimate the relative populations of ferromagnetic and superparamagnetic particles in the sample.  
Let us express the effective magnetic moment of an aggregate of particles as ${\bf m}(t)={\bf m}_0+\delta{\bf m}(t)$, where ${\bf m}_0$ is associated to the ferromagnetic particles and $\delta{\bf m}(t)$ to the fast-fluctuating superparamagnetic particles. The magnetic field produced by the aggregate writes ${\bf B}(t)={\bf B}_0+\delta{\bf B}(t)$. Because of spatial averaging over many NV centre locations within the nanodiamond, it is a good approximation to write
\begin{equation} \label{EqRatio}
\frac{\sqrt{\langle \delta m(t)^2\rangle}}{m_0}\approx\frac{\sqrt{\langle\overline{\delta B(t)^2}\rangle}}{\overline{B_0}}.
\end{equation}
Since the magnetic moment ${\bf m}$ is proportional to the volume of the particles, the ratio (\ref{EqRatio}) can be interpreted as the ratio in volume between superparamagnetic and ferromagnetic particles, which is of the order of 10\% according to the data of main-text Fig. 2. This ratio is consistent with the measured particle size distribution of $5-25$ nm and the estimated threshold size of 10 nm for the ferromagnetic/superparamagnetic transition.

\subsection{Supplementary data}

In main-text Fig. 3, spin resonance shift and spin relaxation images of a small aggregate of maghemite particles are shown (AFM image reproduced in Fig. \ref{FigS4}a). These images represent the differential signal ${\cal D}(\nu_1,\nu_2)=[{\cal F}(\nu_2)-{\cal F}(\nu_1)]/{\cal F}(\nu_2)$ and ${\cal D}(\tau_1,\tau_2)=[{\cal F}(\tau_1)-{\cal F}(\tau_2)]/{\cal F}(\tau_1)$, respectively. To quantify more precisely the strength of the magnetic field produced by the sample, we show here full spin resonance spectra (Fig. \ref{FigS4}b) and $T_1$ spin relaxation curves (Fig. \ref{FigS4}c) recorded at two different locations on the sample. Following the same analysis as in sections \ref{secODMR} and \ref{secT1}, we find a maximum value of $\sqrt{\langle\overline{\delta B(t)^2}\rangle}\approx160~\mu{\rm T}$-rms for the fluctuating component, while the static component is $\overline{B_{0}}\approx100~\mu$T. We conclude that the aggregate imaged in main-text Fig. 3 is dominated by superparamagnetic particles.

\begin{figure*}[hbt]
\begin{center}
\includegraphics[width=0.8\textwidth]{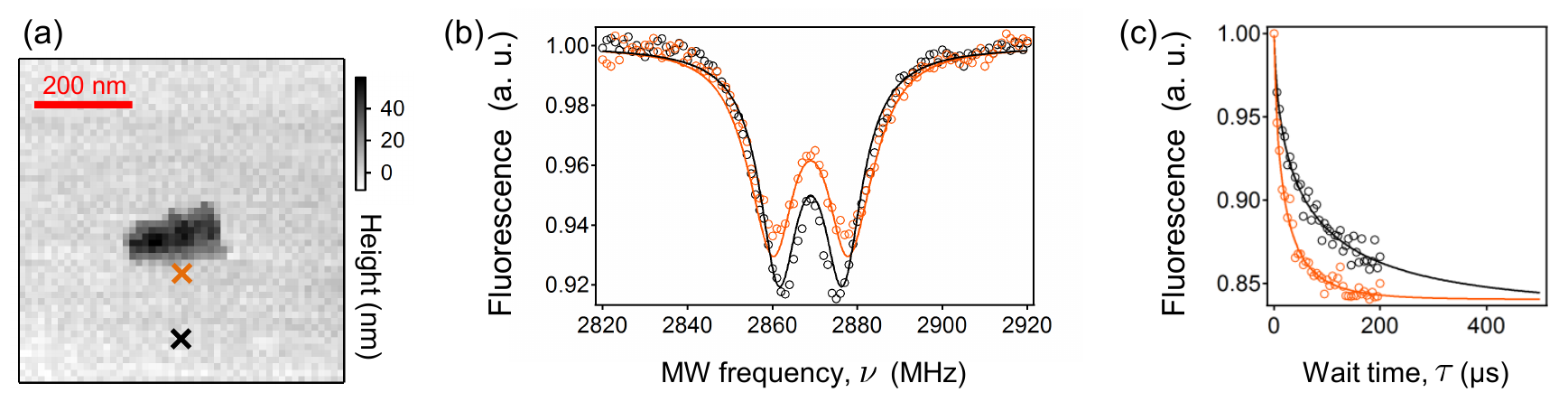}
\caption{(a) AFM image of a small aggregate of maghemite particles, reproduced from main-text Fig. 3a. (b) Optically detected spin resonance spectra recorded with the tip at two different positions, indicated by crosses in (a). The solid lines are data fit as explained in Section \ref{secODMR}. The black curve is a reference spectrum recorded far from the aggregate, while the orange curve close to the aggregate gives $\overline{B_{0}}=0.1\pm0.1$ mT and $\sigma_{B_{0,\parallel}}=0.0\pm0.1$ mT. (c) Spin relaxation curves recorded at the same two positions. The solid lines are data fit to a stretched exponential function, yielding a reference relaxation time $T_1=51.3\pm16.2~\mu$s (black curve) and $T_1=15.8\pm2.3~\mu$s (orange curve).}
\label{FigS4}
\end{center}
\end{figure*}

\subsection{Theoretical estimates of the magnetic field produced by a magnetic particle}

In this section, we compare our measurements of the magnetic field produced by aggregates of maghemite particles to simple theoretical estimates. In the experiment, the nanodiamond has a diameter of $\approx100$ nm and oscillates vertically with an amplitude of $\approx60$ nm. This implies that the nanodiamond centre is located on average at a distance $\bar{h}\approx110$ nm above the sample's surface. Since $\bar{h}$ is significantly larger than the typical diameter $d$ of the maghemite particles ($d=5-25$ nm), we can treat the particles as point-like magnetic dipoles in order to calculate the field probed by the nanodiamond. 

The magnetic moment of a particle of diameter $d$ and magnetisation $M$ has a magnitude $m=\frac{\pi d^3}{6}M$ and is oriented along the easy axis of the particle. The amplitude of the magnetic field right above the particle is then given by 
\begin{equation} \label{EqDipole}
B=\frac{\mu_0 m}{4\pi h^3}(1+3\cos^2\theta_m),
\end{equation}
where $\theta_m$ is the angle between the moment ${\bf m}$ and the vertical direction and $h$ is the distance to the particle. Using $M=300$ kA/m (obtained from Fig. \ref{FigS3b}b), $h=110$ nm and $d=20$ nm, Eq. (\ref{EqDipole}) predicts that the field amplitude varies between 100 $\mu$T for $\theta_m=\pi/2$ and 400 $\mu$T for $\theta_m=0$. Therefore, an aggregate of several particles is expected to produce a field in the mT range, with a static component associated to the ferromagnetic particles and a fluctuating one caused by the superparamagnetic particles. This order of magnitude is in agreement with the experiment, where field amplitudes up to 2.4 mT (static) and 270 $\mu$T-rms (fluctuating) were measured.

\section{Supporting information for the thermal imaging experiments}

\subsection{Sample preparation and characterisation}

The gold nanoparticles used for the experiments reported in main-text Fig. 4 were synthesised following the protocol described in Ref. \cite{Baik2011}. Briefly, we proceed in two steps. First a gold seed solution is prepared, with a diameter $d=13\pm1$ nm. To this end 100 mL of a 1.0 mM aqueous HAuCl$_4$:3H$_2$O solution was added to 100 mL of milli-Q water, which was then boiled. 10 mL of a 38.8 mM aqueous solution of sodium citrate was added and boiled for 30 min. Next, a seed-mediated growth approach was applied in order to synthesize gold particles of desired size. A volume of 4 mL of a 20 mM aqueous HAuCl$_4$:3H$_2$O solution and 0.4 mL volume of a 10 mM aqueous AgNO$_3$ solution were added to 170 mL of milli-Q water. 3 mL of the 13-nm gold seed solultion was added to the solution. 30 mL of 5.3 mM ascorbic acid was then added slowly (0.6 mL/min) with constant stirring. 

\begin{figure*}[b]
\begin{center}
\includegraphics[width=0.4\textwidth]{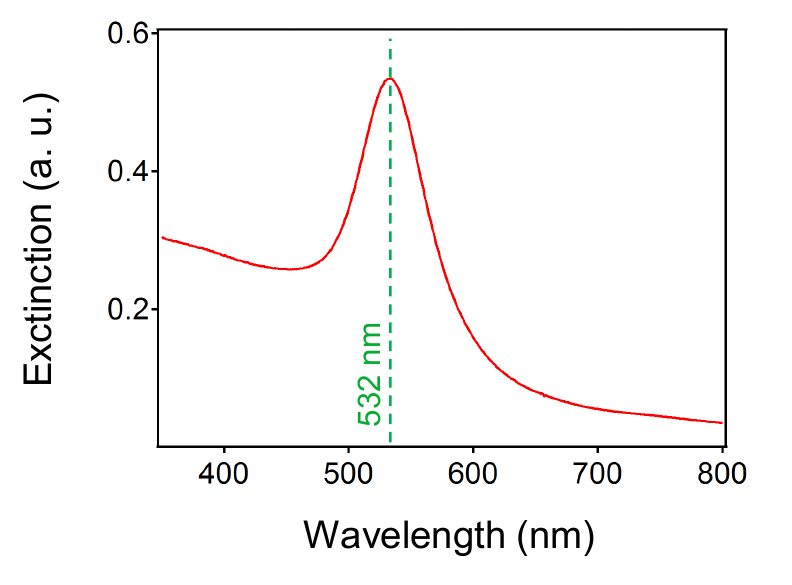}
\caption{Extinction spectrum of the solution of gold nanoparticles recorded prior to deposition on the substrate. The surface plasmon resonance matches well the wavelength of the excitation laser used in the experiment (532 nm).}
\label{FigS5}
\end{center}
\end{figure*}

The resulting solution contains gold particles with a mean diameter of $\approx30$ nm, as measured by dynamic light scattering. Fig. \ref{FigS5} shows an extinction spectrum of the solution. It exhibits a surface plasmon resonance at a wavelength of $\approx530$ nm, which matches well the wavelength of the laser used in the experiment to excite the NV centres. This implies that the same laser will be efficiently absorbed by the gold particles, eventually causing significant heating. A sample of mostly isolated gold particles was prepared by spin casting a drop of the solution on a glass cover slip. An AFM image of the sample is shown in Fig. \ref{FigS6}a. By measuring the height of each particle (Figs. \ref{FigS6}b and \ref{FigS6}c), we find the diameter to vary between 20 and 40 nm. 

\begin{figure*}[t]
\begin{center}
\includegraphics[width=0.8\textwidth]{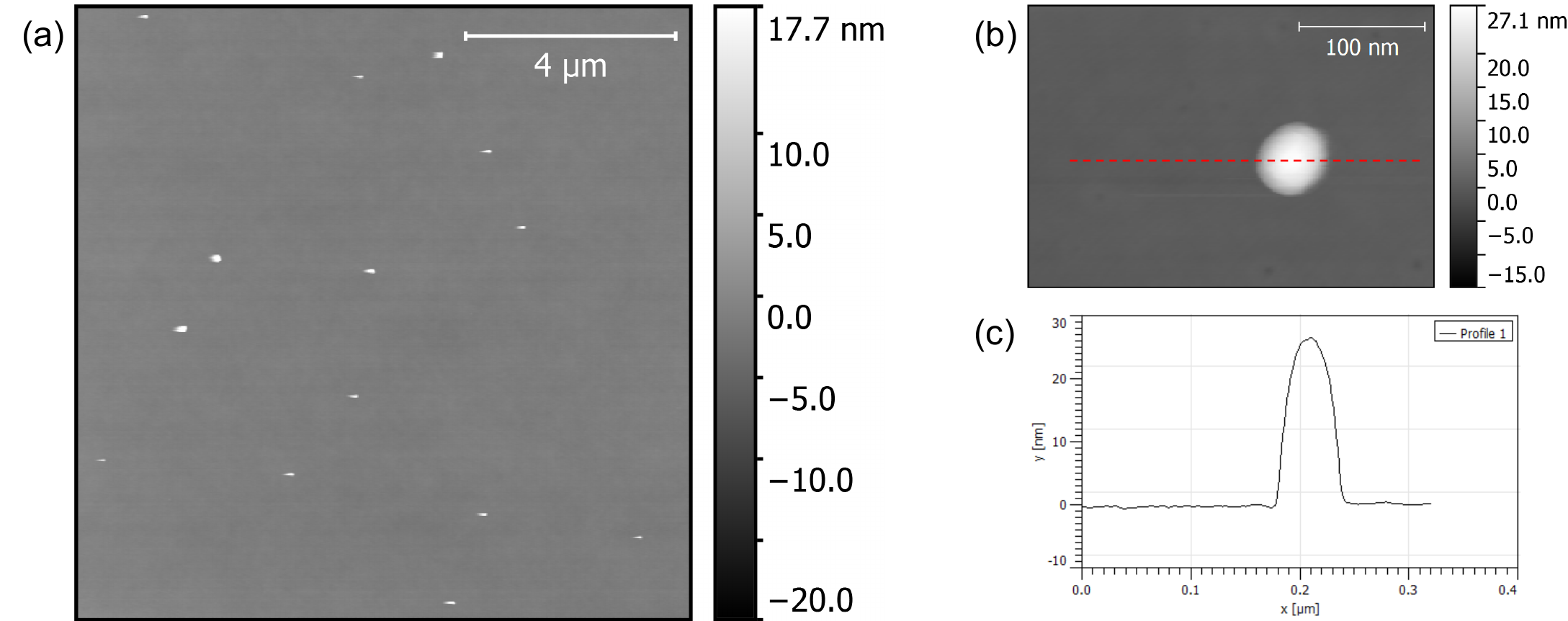}
\caption{(a) AFM image of a sample of gold nanoparticles. (b) Close-up view of a single particle. (c) Line cut taken along the red dashed line in (b).}
\label{FigS6}
\end{center}
\end{figure*}

\subsection{Pre-characterisation of the gold particle investigated}
 
For this experiment, an AFM cantilever equipped with a plateau-shaped tip was used and the nanodiamond was attached to the side wall of the plateau (see Fig. \ref{FigS7}a). As explained below, this enables us to scan the nanodiamond above the gold particle without any direct contact between tip and particle, thus preventing unwanted friction effects. 

Fig. \ref{FigS7}b shows a confocal image recorded prior to the AFM scans. The position of the nanodiamond as well as the plateau tip to which it is attached are indicated. Also shown is the gold particle that was scanned in main-text Fig. 4. Its fluorescence signal is measurable although significantly weaker than that of the nanodiamond. Figs. \ref{FigS7}c and \ref{FigS7}d show the AFM and fluorescence images obtained simultaneously by scanning the sample (i.e., the gold particle) while focusing the laser on the nanodiamond. Since the AFM image reflects the convolution between the particle shape and the tip shape, the particle appears here as a crescent moon due to the plateau being slightly tilted with respect to the substrate's surface. The maximum height observed in the AFM image is indicative of the size of the gold particle, which is $\approx40$ nm (Fig. \ref{FigS7}e). On the other hand, the fluorescence image exhibits a bright spot stemming from the gold particle. Since the laser is focused on the nanodiamond, this spot corresponds to the situation where the nanodiamond is exactly above the gold particle. The absence of overlap between the fluorescent spot and the topography feature of the gold particle implies that the nanodiamond can be scanned around the gold particle with a constant nanodiamond-substrate distance, without direct contact between tip and gold particle. Furthermore, it is sufficient to use the fluorescence signal to centre the scan on the gold particle, without resorting to the topography information.  In main-text Figs. 4b and 4c, we thus scanned the nanodiamond probe around the gold particle while recording spin resonance spectra, allowing us to form a temperature map of the photo-heated gold particle.

\begin{figure*}[htb]
\begin{center}
\includegraphics[width=1\textwidth]{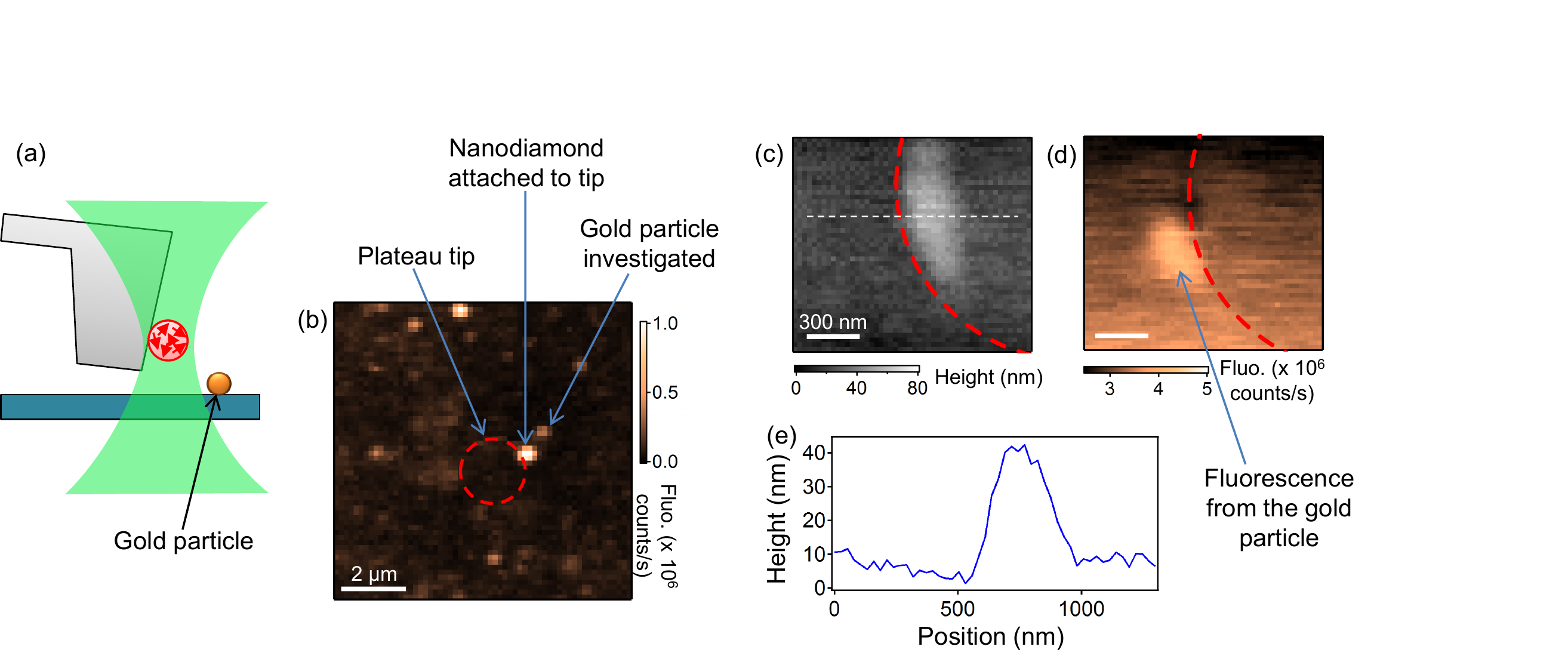}
\caption{(a) Schematic of the experiment showing the nanodiamond attached to the side wall of a plateau-shaped tip. (b) Confocal image obtained by scanning the laser beam relative to sample and AFM tip, with the AFM tip engaged. The red circle depicts the 2-$\mu$m-diameter plateau. (c,d) Topography image (c) and fluorescence image (d) obtained simultaneously by scanning the sample relative to the AFM tip, with the laser being focused on the nanodiamond. The red circle delimits the topographic feature that would results from the plateau being perfectly parallel to the substrate's surface. (e) Line cut taken along the white dashed line in (c), indicating a particle size of $\approx40$ nm.}
\label{FigS7}
\end{center}
\end{figure*} 

\subsection{Acquisition of the temperature map}

The images reported in main-text Figs. 4b and 4d were obtained by scanning the sample stage while operating the AFM tip in AC mode. A drop of milli-Q water was added onto the substrate prior to mounting the AFM head. We made sure that the AFM cantilever was fully immersed in water by monitoring its resonance frequency, which dropped from $\approx300$ kHz in air to $\approx120$ kHz in water. The oscillation amplitude was estimated from force curves to be $\approx 10$ nm in scanning conditions. The $XYZ$ position of the objective lens was adjusted after each line of the scan in order to maximise the NV fluorescence signal. In main-text Fig. 4d, a spin resonance spectrum composed of 25 frequency bins was acquired at each pixel, some of the obtained spectra being shown in main-text Fig. 4c. The time per bin is 60 ms, resulting in a pixel time of 1.5 s. Both laser and microwave were applied continuously during the acquisition, with an incident laser power of $P_L\approx250~\mu$W.

Each individual spectrum was fitted using Eq. (\ref{Eq1}). The obtained resonance frequencies $\nu_\pm$ allow us to infer the splitting parameter $D=\frac{\nu_++\nu_-}{2}$. The temperature increase $\Delta T$ plotted in main-text Fig. 4d was calculated according to
\begin{equation} \label{eqTemp}
\Delta T=\frac{D-D_0}{-75~{\rm kHz/K}},
\end{equation}
where the numerical factor is the linear conversion factor measured by Acosta {\it et al.} near room temperature \cite{Acosta2010}. The offset $D_0$ is the splitting parameter measured in the absence of laser heating. To obtain $D_0$, we recorded spin resonance spectra as a function of the laser power $P_L$, with the tip engaged on the sample far from any gold particle. Fig. \ref{FigS8} shows the parameter $D$ extracted from these spectra as a function of $P_L$. A linear fit provides us with $D_0=D(P_L=0)=2868.9\pm0.2$ MHz and a slope of $-2.8\pm0.4$ kHz/$\mu$W. This slope indicates the existence of a laser-induced heating effect even in the absence of the gold particle. This most likely occurs because some laser light is absorbed by the silicon AFM tip, thereby heating the nanodiamond attached to it. With the laser power of $P_L\approx250~\mu$W used in the experiment (main-text Fig. 4c), this intrinsic heating is $\Delta T\approx9$ K. However, this temperature increase is significantly smaller than that caused by laser-induced heating of the gold particle, as clearly seen in main-text Fig. 4e.

\begin{figure*}[t]
\begin{center}
\includegraphics[width=0.4\textwidth]{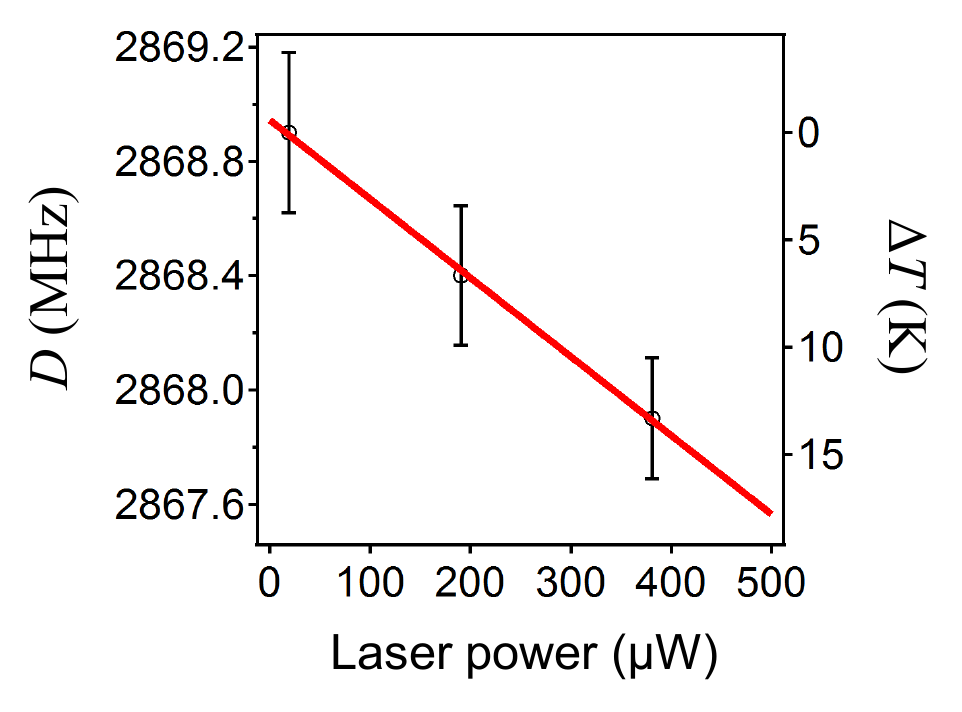}
\caption{Splitting parameter $D$ of the NV ensemble as a function of the incident laser power $P_L$. The measurement conditions are the same as in main-text Fig. 4c, i.e., the nanodiamond is attached to the plateau tip, the tip is engaged far from any gold particle and the laser is focused on the nanodiamond. Shown on the right-hand side is the corresponding temperature increase $\Delta T$ deduced from Eq. (\ref{eqTemp}).}
\label{FigS8}
\end{center}
\end{figure*} 

\subsection{Derivation of the analytical formula used to fit the measured temperature profile}

In this section we derive the analytical formula given in the main text (Eq. (1)), which represents the nanodiamond temperature $\Delta T_d(x_g)$ as a function of the position $x_g$ of the gold particle relative to the nanodiamond and laser beam, as experimentally measured in main-text Fig. 4e. When the gold particle is moved horizontally to a new position $x_g$, two effects contribute to change the temperature of the nanodiamond probe: (i) the gold particle receive a different amount of heating power because of the intensity profile of the stimulating laser beam, and (ii) for a given heating power the temperature at the diamond location is changed because the distance to the gold particle is changed. Hereafter these two independent effects are modelled in order to obtain an analytical formula.  

\paragraph{Heating power vs. position}

\begin{figure*}[b]
\begin{center}
\includegraphics[width=0.4\textwidth]{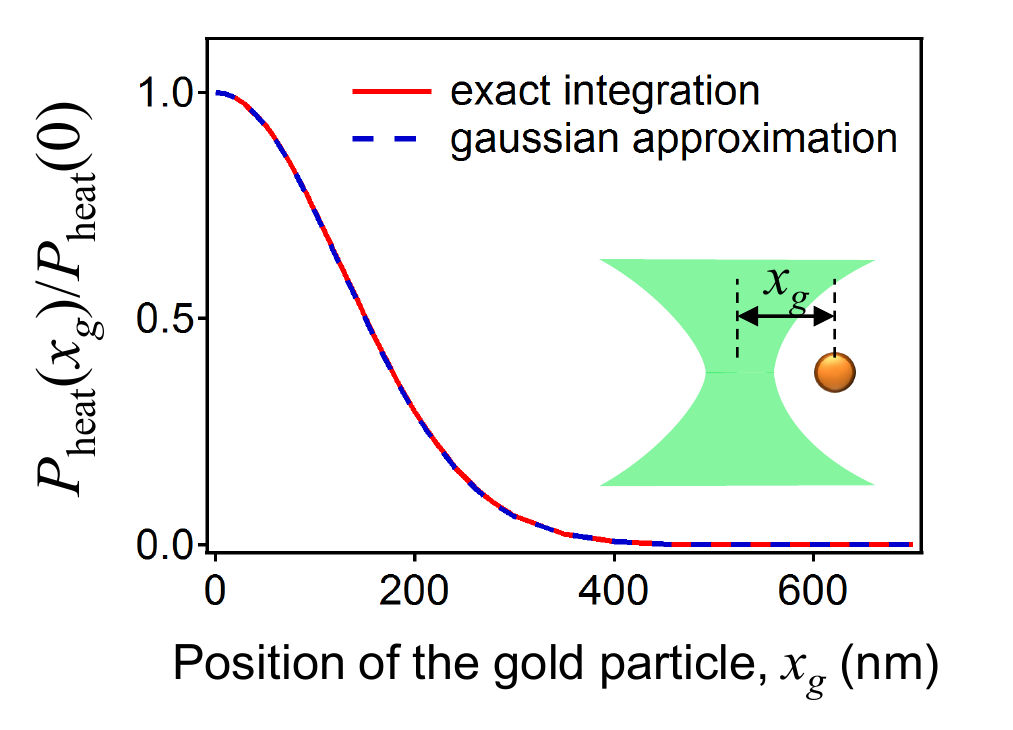}
\caption{Normalised power absorbed by a gold nanosphere as a function of the relative position of the exciting beam. The radius of the sphere is 20 nm and the FWHM of the Gaussian beam is 300 nm. }
\label{FigS9}
\end{center}
\end{figure*} 

The transverse profile of the laser beam is modelled by a 2D Gaussian distribution with a full width at half maximum (FWHM) of 300 nm determined experimentally through confocal imaging of a single NV centre. The power density is denoted as $p_L({\bf r})$ with ${\bf r}=(x,y)$ being the transverse position, and reads
\begin{equation}
p_L({\bf r})=\frac{P_L}{2\pi\sigma_L^2}\exp\left(-\frac{r^2}{2\sigma_L^2}\right),
\end{equation}
where $\sigma_L\sqrt{8\ln2}=300$ nm and $P_L=\iint p_L({\bf r}){\rm d}^2{\bf r}=250~\mu{\rm W}$ is the total laser power. We consider a spherical gold particle of radius $R_g$. The power absorbed by the particle can be written as
\begin{equation} 
P_{\rm abs}(x_g)=\sigma_{\rm abs}I_L(x_g),
\end{equation} 
where $\sigma_{\rm abs}$ is the effective absorption cross section of the particle and $I_L(x_g)$ is the laser intensity averaged over the cross-sectional area of the particle, that is,
 \begin{equation} \label{eqinteg}
I_L(x_g)=\frac{1}{\sigma_{\rm geom}}\iint_{{\cal A}(x_g)}p_L({\bf r}){\rm d}^2{\bf r},
\end{equation} 
with ${\cal A}(x_g)$ being the geometric cross section of the particle of total area $\sigma_{\rm geom}=\pi R_g^2$. For a gold sphere of radius 20 nm stimulated at a wavelength of 532 nm, Mie theory \cite{BohrenHuffman} predicts a ratio $\frac{\sigma{\rm abs}}{\sigma{\rm geom}}\approx3.0$, which yields a maximum absorbed power of $P_{\rm abs}(0)\approx9.0~\mu{\rm W}$. Fig. \ref{FigS9} shows the relative absorbed power $P_{\rm abs}(x_g)/P_{\rm abs}(0)$ calculated from Eq. (\ref{eqinteg}), as well as from the approximation by a Gaussian function
\begin{equation} \label{eqgauss}
\frac{P_{\rm abs}(x_g)}{P_{\rm abs}(0)}\approx \exp\left({-\frac{x_g^2}{2\sigma_L^2}}\right),
\end{equation} 
with $\sigma_L\sqrt{8\ln2}=300$ nm being the FWHM of the laser beam. In the following, we will use the Gaussian approximation (\ref{eqgauss}), which is in excellent agreement with the exact integration. Moreover, we will assume that all the absorbed power is converted into heat, that is, $P_{\rm abs}(x_g)$ represents the heating power generated by the gold particle for a given position $x_g$.

\paragraph{Temperature distribution}

We now analyse the temperature distribution outside the gold particle for a given heating power $P_{\rm abs}$. Some of the assumptions and approximations made in this paragraph will be justified in the next section by comparison with numerical simulations. In the steady-state regime, the thermal diffusion equation is
\begin{equation} \label{eqDiffusion}
\nabla\cdot[\kappa({\bf r})\nabla T({\bf r})]=-p_{\rm heat}({\bf r}),
\end{equation}
where $\kappa({\bf r})$ is the position-dependent thermal conductivity and $p_{\rm heat}({\bf r})$ is the power density of the heat source. Here, the only source of heat is the gold particle with a total power $\iiint p_{\rm heat}({\bf r}){\rm d}^3{\bf r}=P_{\rm abs}$. The particle is modelled as a sphere of radius $R_g$ in a homogeneous medium of thermal conductivity $\kappa_{\rm eff}$. Because the thermal conductivity of gold is much larger than that of the surrounding medium, the temperature in the gold particle does not depend on the exact distribution $p_{\rm heat}({\bf r})$ and can be considered as uniform \cite{Baffou2010}, with a value
\begin{equation} \label{eqTg}
\Delta T_g=\frac{P_{\rm abs}}{4\pi R_g\kappa_{\rm eff}}.
\end{equation}
Outside the gold sphere, the temperature distribution reads \cite{Baffou2010}
\begin{equation} \label{eqT}
\Delta T(r)\approx\Delta T_g\frac{R_g}{r},
\end{equation} 
where $r>R_g$ is the distance from the gold sphere centre. In the experiment, the gold particle is moved horizontally (position $x_g$) while the nanodiamond probe is standing still at a height $h_{gd}$, defined as the centre-to-centre distance between the gold sphere and the nanodiamond when the two particles are vertically aligned (i.e. $x_g=0$). We assume that the diamond temperature $\Delta T_d$ can be approximated by the value calculated at the centre of the diamond volume in its absence, that is, 
\begin{equation} \label{Td1}
\Delta T_d(x_g)\approx\Delta T(r=\sqrt{x_g^2+h_{gd}^2})\approx\Delta T_g(x_g)\frac{R_g}{\sqrt{x_g^2+h_{gd}^2}}.
\end{equation} 
Using Eqs. (\ref{eqgauss}) and (\ref{eqTg}), one can write
\begin{equation} \label{eqgauss2}
T_g(x_g)\approx T_g(0)\exp\left({-\frac{x_g^2}{2\sigma_L^2}}\right),
\end{equation} 
where $T_g(0)=P_{\rm abs}(0)/4\pi R_g\kappa_{\rm eff}$. Inserting Eq. (\ref{eqgauss2}) into Eq. (\ref{Td1}), one finally obtains
\begin{equation} \label{eqDTfinal}
\Delta T_d(x_g)\approx\Delta T_g(0)\exp\left(-\frac{x_g^2}{2\sigma_L^2}\right)\frac{R_g}{\sqrt{x_g^2+h_{gd}^2}},
\end{equation} 
which is Eq. (1) of the main text. The temperature of the gold particle at $x_g=0$ ($T_g(0)$) and the height of the diamond probe ($h_{gd}$) are used as fit parameters, while the other parameters are taken to be $R_g=20$ nm and $\sigma_L\sqrt{8\ln2}=300$ nm. 

\paragraph{Comparison with numerical simulations}

COMSOL Multiphysics (v5.1) was used to simulate the temperature distribution of a system comprising  a gold nanosphere on silica glass and a diamond nanosphere suspended in water (Fig. \ref{FigS10}a). The simulations were performed on a diamond nanosphere (radius 50 nm) suspended in a hemisphere of water (radius 50 $\mu$m) 60 nm above a cylinder of silica glass (radius and height 50 $\mu$m), and a gold nanosphere (radius 20 nm) placed on the surface of the silica glass (Fig. \ref{FigS10}b). These dimensions were chosen to be representative of the experimental situation. The gold nanosphere is heated with a constant power $P_{\rm abs}(x_g)$ following Eq. (\ref{eqgauss}). As a test value, we use a heating power of $P_{\rm abs}(0)=36~\mu$W. A boundary condition of constant temperature at the ambient temperature of 296.15 K was set to replicate an infinite medium of water and silica glass. The calculation domains were meshed with a physics controlled extremely fine triangular mesh to ensure correct geometric treatment of the gold and diamond nanospheres and to increase accuracy of heat transfer. The program then solves the thermal diffusion equation (\ref{eqDiffusion}) in the steady-state regime. We note that the steady state is reached within a $\mu$s time scale \cite{Baffou2010}, which is much faster that our measurement time.

\begin{figure*}[t]
\begin{center}
\includegraphics[width=0.7\textwidth]{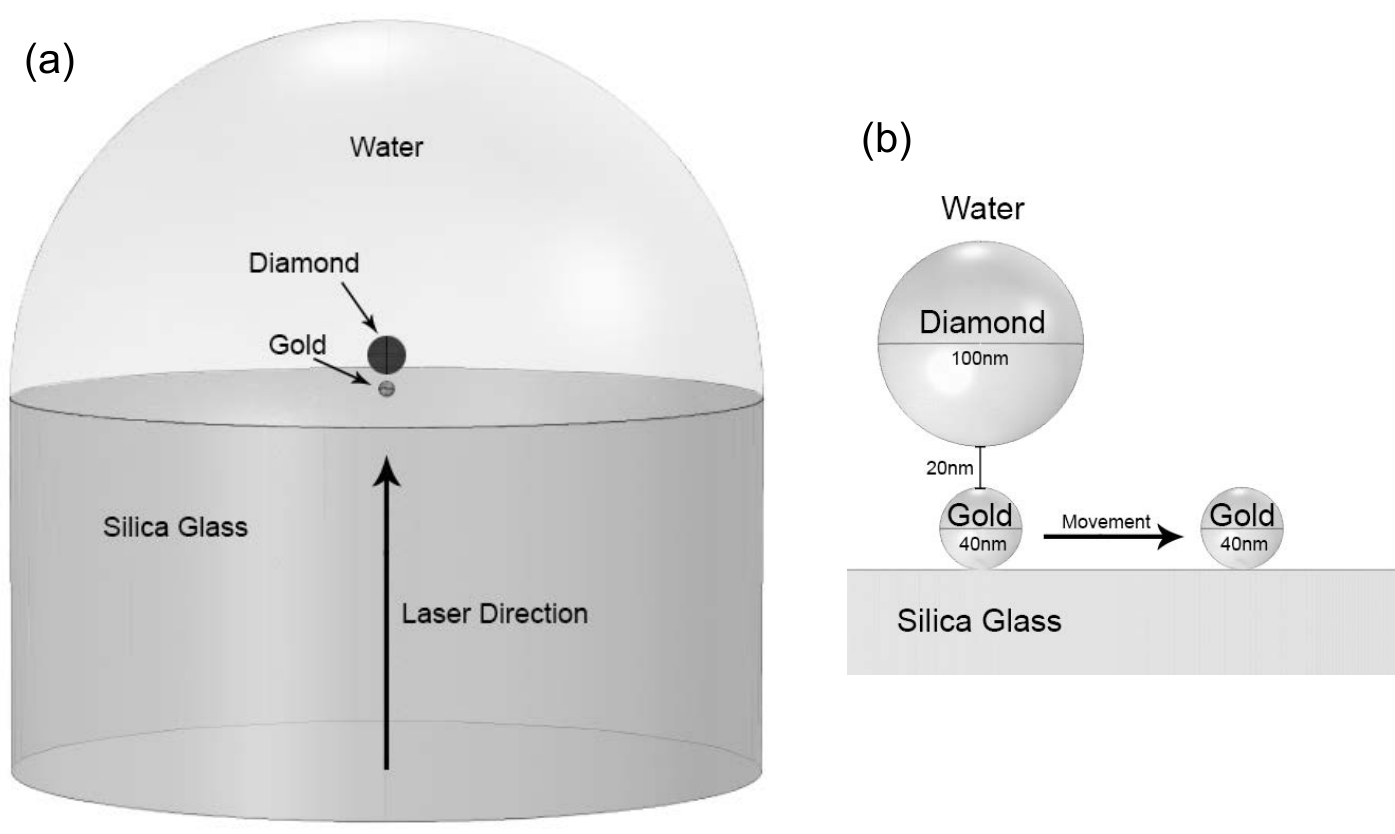}
\caption{(a) Geometry of the domains considered in the thermal simulation. (b) Close-up view showing the dimensions. The simulation is run for different positions of the gold nanosphere. For each position $x_g$, a constant heating power $P_{\rm heat}(x_g)$ is set.}
\label{FigS10}
\end{center}
\end{figure*}  

\begin{figure*}[b]
\begin{center}
\includegraphics[width=1\textwidth]{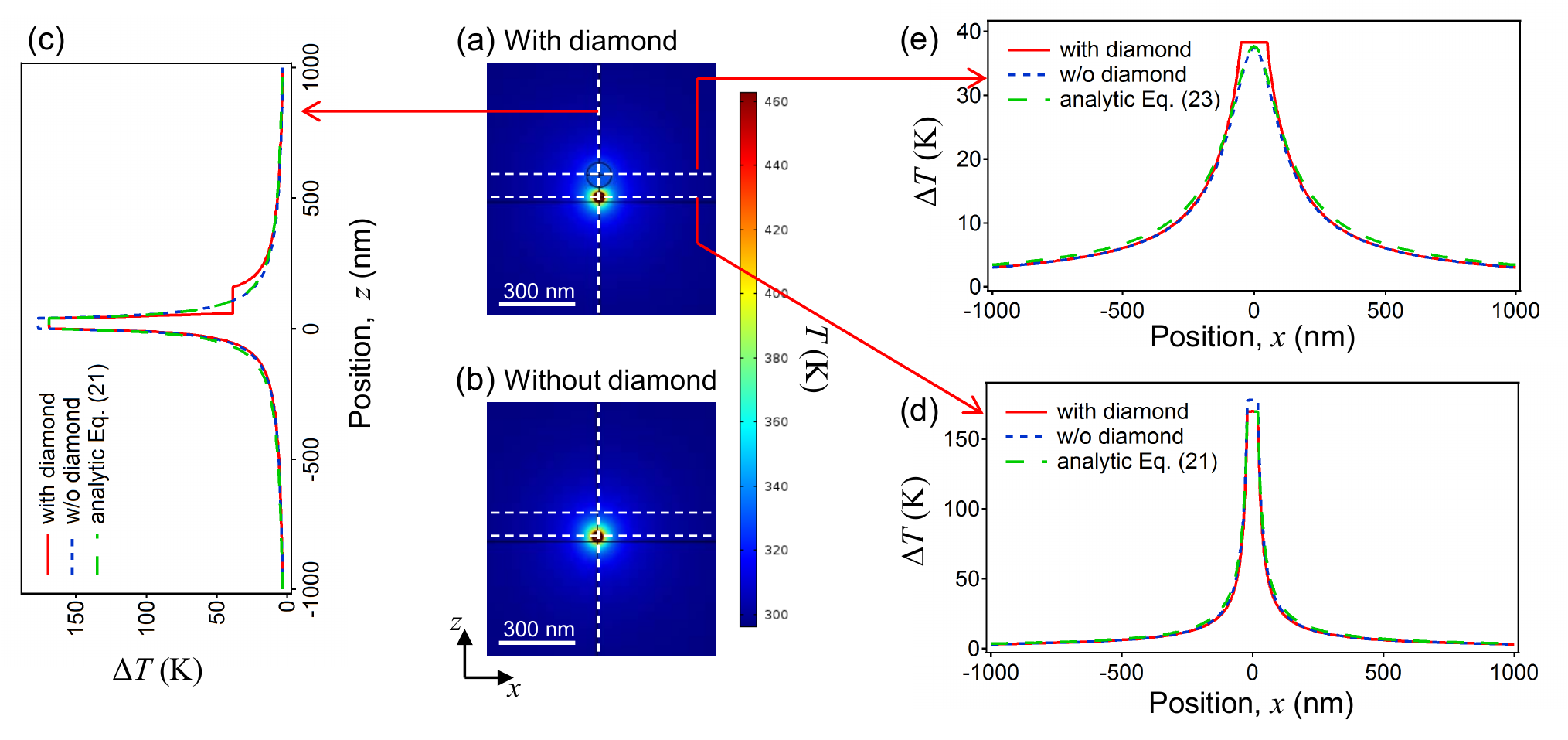}
\caption{(a) Calculated temperature distribution in the $xz$ plane with $x_g=0$ and a heating power $P_{\rm heat}=36~\mu$W. (b) Same as (a) but with the diamond replaced with water. (c-e) Line cuts taken along the white dashed lines shown in (a,b).}
\label{FigS11}
\end{center}
\end{figure*}  

We first consider the case where the gold nanosphere is aligned with the nanodiamond and the laser beam, i.e. $x_g=0$. The resulting temperature map in the $xz$ plane is shown in Fig. \ref{FigS11}a. Also shown for comparison is the case where the diamond is replaced with water (Fig. \ref{FigS11}b). Figs. \ref{FigS11}c-e show line cuts taken along the vertical direction (c), and along the horizontal direction across the gold sphere (d) and across the diamond sphere (e). In the absence of the diamond, the temperature increase inside the gold sphere is nearly uniform and reaches $\Delta T_g\approx177$ K. Outside the gold sphere, the temperature decays quasi isotropically because the thermal conductivity of glass and water are similar. The presence of the diamond sphere affects only marginally the general temperature distribution around the gold particle: because of the high conductivity of diamond, $\Delta T_g$ is slightly decreased by $\approx7$ K, and the temperature in the diamond sphere $\Delta T_d\approx36$ K is nearly uniform, approximately equal to the temperature at the centre of the same volume when replaced with water (see Figs. \ref{FigS11}c and \ref{FigS11}e).

These numerical simulations can be compared with the analytical expressions derived above. According to Eq. (\ref{eqTg}), the temperature of the gold particle in the presence of the diamond, $\Delta T_g=170$ K, corresponds to an effective thermal conductivity of the system (glass + water + diamond nanosphere) of $\kappa_{\rm eff}=0.84$ W$\cdot$m$^{-1}\cdot$K$^{-1}$. This is close to the conductivity of glass and water ($0.8$ W$\cdot$m$^{-1}\cdot$K$^{-1}$ and $0.6$ W$\cdot$m$^{-1}\cdot$K$^{-1}$, respectively. In Figs. \ref{FigS11}c-e, we compare the numerical simulations with the predictions of Eq. (\ref{eqT}) using $\Delta T_g=170$ K, which are found to be in excellent agreement. We can also run the simulation for various positions $x_g$ of the gold particle, taking into account the position-dependent heating power as given by Eq. (\ref{eqgauss}). The resulting curve $T_d(x_g)$ is in very good agreement with the analytic formula (\ref{eqDTfinal}), as illustrated in Fig. \ref{FigS12}. In main-text Fig. 4e, the data were therefore fitted to Eq. (\ref{eqDTfinal}) to extract the temperature of the gold particle $\Delta T_g$ as well as the mean distance $h_{gd}$. 

\begin{figure*}[t]
\begin{center}
\includegraphics[width=0.5\textwidth]{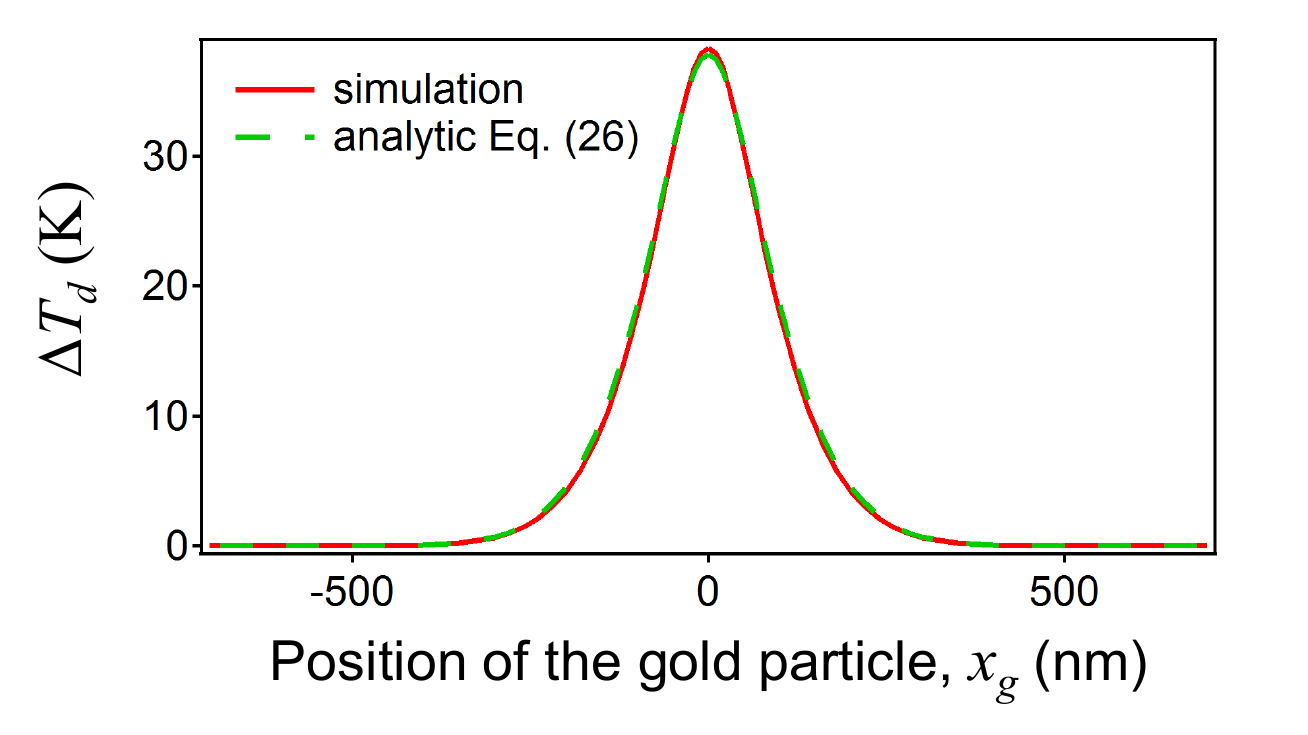}
\caption{Temperature increase of the diamond nanosphere as a function of the position of the gold nanosphere, as obtained from the numerical simulation (red line) or from Eq. (\ref{eqDTfinal}).}
\label{FigS12}
\end{center}
\end{figure*}  

\vspace{1 cm}
\end{widetext}

\end{document}